\documentclass[preprint,aps,eqsecnum,showkeys,draft]{revtex4}
\usepackage{amsfonts}

\begin{document}


\renewcommand{\baselinestretch}{1.7} 

\def\eqref#1{(\ref{#1})}
\def\eqrefs#1#2{(\ref{#1}) and~(\ref{#2})}
\def\eqsref#1#2{(\ref{#1}) to~(\ref{#2})}
\def\sysref#1#2{(\ref{#1})--(\ref{#2})}

\def\Eqref#1{Eq.~(\ref{#1})}
\def\Eqrefs#1#2{Eqs.~(\ref{#1}) and~(\ref{#2})}
\def\Eqsref#1#2{Eqs.~(\ref{#1}) to~(\ref{#2})}

\def\EQ #1\doneEQ{\begin{equation} #1 
\end{equation}}
\def\EQs #1\doneEQs{\begin{eqnarray} #1 
\end{eqnarray}}

\def\secref#1{Section~\ref{#1}}
\def\secrefs#1#2{Sections~\ref{#1} and~\ref{#2}}
\def\secsref#1#2{Sections~\ref{#1} to~\ref{#2}}

\def\Ref#1{Ref.\cite{#1}}
\def\Refs#1{Refs.\cite{#1}}


\def\eqtext#1{\hbox{\rm{#1}}}

\def\mstrut{\mathstrut}
\def\hp#1{\hphantom{#1}}
\def\smallfrac#1#2{{\textstyle{{#1}\over{#2}}}}

\def\mixedindices#1#2{{\mstrut}^{\mstrut #1}_{\mstrut #2}}
\def\downindex#1{{\mstrut}^{\mstrut}_{\mstrut #1}}
\def\upindex#1{{\mstrut}_{\mstrut}^{\mstrut #1}}
\def\downupindices#1#2{{\mstrut}_{\mstrut #1}^{\hp{#1}\mstrut #2}}
\def\updownindices#1#2{{\mstrut}^{\mstrut #1}_{\hp{#1}\mstrut #2}}

\def\der#1{\partial\downindex{#1}}
\def\coder#1{\partial\upindex{#1}}
\def\covder#1{\nabla\downindex{#1}}
\def\covcoder#1{\nabla\upindex{#1}}
\def\perpcovder#1{\nabla\mixedindices{\perp}{#1}}

\def\Sder#1{{\cal D}\downindex{#1}}
\def\coSder#1{{\cal D}\upindex{#1}}
\def\covSder#1{\nabla\mixedindices{S}{#1}}
\def\covcoSder#1{\nabla\upindex{S #1}}
\def\perpSder#1{{\cal D}\mixedindices{\perp}{#1}}
\def\perpcoSder#1{{\cal D}\upindex{\perp #1}}

\def\D#1{D\downindex{#1}}
\def\coD#1{D\upindex{#1}}

\def\slD#1{
\setbox1=\hbox{$D$}
{\hbox to\wd1{$\box1\mkern-11mu\mathchar'57$\hfill}}_{#1} }
\def\slSder#1{
\setbox1=\hbox{$\cal D$}
{\hbox to\wd1{$\box1\mkern-11mu\mathchar'57$\hfill}}_{#1} }
\def\ethD#1{
\setbox1=\hbox{$\cal D$}
\setbox2=\hbox{$\mathchar'55$}
{\hbox to\wd1{$\box1 \mkern-12mu\raise2pt\box2$\hfill}}_{#1} }
\def\slethD#1{
\setbox1=\hbox{$\cal D$}
\setbox2=\hbox{{$\mathchar'55$}}
{\hbox to\wd1
{$\box1 \mkern-11mu\mathchar'57\mkern-10mu\raise2pt\box2$\hfill}}_{#1} }

\def\div#1{{\rm div}_{#1}}

\def\g#1#2{g\downupindices{#1}{#2}}
\def\metric#1#2{\sigma\downupindices{#1}{#2}}
\def\perpmetric#1#2{\sigma\mixedindices{\perp}{#1}\upindex{#2}}
\def\flat#1#2{\eta\downupindices{#1}{#2}}
\def\vol#1#2{\epsilon\downupindices{#1}{#2}}
\def\duvol#1#2{{*\epsilon}\downupindices{#1}{#2}}
\def\perpvol#1#2{\epsilon\mixedindices{\perp}{#1}\upindex{#2}}
\def\coordvol#1#2{\varepsilon\downupindices{#1}{#2}}
\def\e#1#2{e_{#1}\upindex{#2}}
\def\inve#1#2{e^{#1}\downindex{#2}}
\def\vole#1#2{{\ast(e\wedge e)}\downupindices{#1}{#2}}

\def\h#1#2{h\downupindices{#1}{#2}}
\def\K#1#2{K\downupindices{#1}{#2}}
\def\trK{K}
\def\R#1#2{R\downupindices{#1}{#2}(\h{}{})}
\def\a#1#2{a\downupindices{#1}{#2}}

\def\x#1#2{x\mixedindices{#1}{#2}}
\def\id#1#2{\delta\mixedindices{#2}{#1}}

\def\curv#1#2{R\downupindices{#1}{#2}}
\def\scurv{R}
\def\ricci{Ric}
\def\conx#1#2{\Gamma\downupindices{#1}{#2}}
\def\T#1#2{T\downupindices{#1}{#2}}

\def\curvS#1#2{{\cal R}\downupindices{#1}{#2}}
\def\scurvS{{\cal R}}
\def\k#1#2{k\mixedindices{#2}{#1}}
\def\trk{\kappa}
\def\shear#1#2{{\cal K}\mixedindices{#2}{#1}}
\def\kperp#1#2{k_\perp\mixedindices{#2}{#1}}
\def\trkperp{\trk_\perp}
\def\norScurv#1#2{{\cal R}\mixedindices{\perp}{#1}\upindex{#2}}
\def\norscurvS{{\cal R}^\perp}

\def\perpcurv#1#2{R\mixedindices{\perp}{#1}\upindex{#2}}
\def\perpconx#1#2{{\cal J}\mixedindices{#1}{#2}}

\def\n{t}
\def\nvec#1#2{\n\mixedindices{#1}{#2}}
\def\u{u}
\def\uvec#1#2{\u\mixedindices{#1}{#2}}
\def\m{m}
\def\cm{\bar m}

\def\outnull{\ell}
\def\innull{n}

\def\frame#1#2{\vartheta\mixedindices{#2}{#1}}
\def\nframe#1#2{\vartheta\mixedindices{#2}{#1}}
\def\outnframe#1#2{\vartheta\mixedindices{+#1}{#2}}
\def\innframe#1#2{\vartheta\mixedindices{-#1}{#2}}
\def\volframe#1#2{{\ast(\vartheta\wedge\vartheta)}\downupindices{#1}{#2}}

\def\indS#1{{\underline #1}}
\def\indH#1{{\hat #1}}

\def\mcurv{H}
\def\perpmcurv{H_\perp}
\def\twist{\varpi}
\def\perptwist{\varpi^\perp}
\def\P{P}
\def\absmcurv{|\mcurv|}
\def\absperpmcurv{|\perpmcurv|}
\def\absP{|\P{}{}|}
\def\sqP{\P{}{}{}^2}
\def\unitP{\hat \P{}{}}
\def\unitmcurv{\hat H}
\def\unitperpmcurv{{\hat H}_\perp}

\def\abseuclmcurv{|\mcurv|_{\rm flat}}
\def\abseuclP{|\P|_{\rm flat}}
\def\absembedmcurv{|\mcurv|_{\rm embed}}
\def\absembedP{|\P|_{\rm embed}}

\def\maxH{\mcurv_{\rm max}}
\def\minH{\mcurv_{\rm min}}

\def\Pvec#1#2{\P\mixedindices{#1}{#2}}
\def\Hvec#1#2{\mcurv\mixedindices{#1}{#2}}
\def\perpHvec#1#2{\mcurv\mixedindices{#1}{\perp #2}}
\def\wvec#1#2{\twist\mixedindices{#1}{#2}}
\def\perpwvec#1#2{\perptwist\mixedindices{#1}{#2}}

\def\mirr{M_{\rm irr}}

\def\TS{T(S)}
\def\TperpS{\TS^\perp}
\def\coTS{T^*(S)}
\def\coTperpS{\coTS^\perp}
\def\TM{T(M)}
\def\coTM{T^*(M)}
\def\ThM{T(\Sigma)}
\def\coThM{T^*(\Sigma)}
\def\Tsurf{T(\partial\Sigma)}
\def\Tperpsurf{\Tsurf^\perp}

\def\hook{\rfloor}
\def\Lie#1{\pounds_{#1}}
\def\unitop{\openone}
\def\proj#1#2{{\cal P}^{#1}_{#2}}

\def\tr{{\rm tr}}
\def\ad{\dagger}

\def\i{{\rm i}}

\def\d#1{{\rm d}{#1}}

\def\pow{p}
\def\fact{{\mit\Gamma}}

\def\flow{\xi}
\def\lapse{N}
\def\shift#1#2{N_\parallel\mixedindices{#1}{#2}}
\def\flowder#1{\partial{#1}/\partial\flow}
\def\flowvec#1#2{\xi\mixedindices{#1}{#2}}

\def\L{{\cal L}}
\def\E#1{{\cal E}\downindex{#1}}

\def\H#1{H_{#1}}
\def\Hdens{{\cal H}}
\def\refHdens#1{{\cal H}^{\rm ref}_{#1}}

\def\embed{|_{\rm flat}}
\def\eucl{|_{\rm Eucl.}}
\def\curvembed{|_{\rm embed}}

\def\qlE#1{E^{\rm #1}}
\def\qlP#1{P^{\rm #1}}
\def\qlJ#1{J^{\rm #1}}
\def\qlabsE#1{{\bar E}{}^{\rm #1}}
\def\qlI{I}

\def\kv{\zeta}


\def\const{{\rm const}}
\def\Rnum{{\mathbb R}}
\def\cc{{\rm c.c.}}
\def\Re{{\rm Re}}

\def\genus{{\rm g}}

\def\ie/{i.e.}
\def\etc/{etc.}

\title{
Mean curvature flow and quasilocal mass for two-surfaces
in Hamiltonian General Relativity 
}

\author{Stephen C. Anco}
\email{sanco@brocku.ca}
\affiliation{%
Department of Mathematics, Brock University, St Catharines,
Ontario, L2S 3A1, Canada
}

\date{\today}

\begin{abstract} 
A family of quasilocal mass definitions that includes as special cases 
the Hawking mass and the Brown-York ``rest mass'' energy 
is derived for spacelike 2-surfaces in spacetime. 
The definitions involve an integral of powers of 
the norm of the spacetime mean curvature vector of the 2-surface, 
whose properties are connected with apparent horizons.
In particular, for any spacelike 2-surface, 
the direction of mean curvature is orthogonal (dual in the normal space) 
to a unique normal direction in which the 2-surface has 
vanishing expansion in spacetime. 
The quasilocal mass definitions are obtained by an analysis of 
boundary terms arising in the gravitational ADM Hamiltonian 
on hypersurfaces with a spacelike 2-surface boundary,
using a geometric time-flow chosen proportional to 
the dualized mean curvature vector field at the boundary surface. 
A similar analysis is made choosing a geometric rotational flow 
given in terms of the twist covector of 
the dual pair of mean curvature vector fields,
which leads to a family of quasilocal angular momentum definitions
involving the squared norm of the twist. 
The large sphere limit of these definitions is shown to yield
the ADM mass and angular momentum in asymptotically flat spacetimes,
while at apparent horizons a quasilocal version of the Gibbons-Penrose
inequality is derived. 
Finally, some results concerning positivity 
are proved for the quasilocal masses,
motivated by consideration of spacelike mean curvature flow of
2-surfaces in spacetime. 
\end{abstract}


\maketitle

\section{ Introduction }

There has been much interest recently 
in the Hamiltonian (initial-value) formulation of 
General Relativity with spatial boundaries,
especially as motivated by work on numerical solution of the Cauchy problem.
This paper is addressed to 
a study of the covariant gravitational Hamiltonian
purely as a variational principle for the Einstein field equations 
in 3+1 dynamical form on hypersurfaces with a spatial boundary 2-surface. 
A main goal will be to derive from the Hamiltonian boundary terms 
a good geometrical definition of quasilocal energy and angular momentum 
for spacelike 2-surfaces in spacetime. 
The results may help shed light on understanding aspects of 
how to set up physically meaningful and mathematically well-posed 
spatial boundary conditions for solving the Einstein equations. 

Using a covariant spacetime formalism, 
in previous work \Refs{paperI,paperII,errata}
an analysis was made of boundary terms that arise 
in the total gravitational Hamiltonian 
with the assumption of 
a fixed (non-dynamical) spatial boundary $B$ in spacetime.
Such a boundary corresponds to 
fixing a time-flow vector field $\flowvec{\mu}{}$
taken to be independent of the gravitational initial data
on a hypersurface $\Sigma$ with a spacelike boundary 2-surface $S$,
and Lie dragging $S$ along $\flowvec{\mu}{}$ 
to produce the boundary hypersurface $B$. 
Starting from the ADM Hamiltonian in covariant form,
the analysis showed that the mathematically allowed boundary terms
and corresponding boundary conditions on the gravitational field variables
required for the total Hamiltonian to be a well defined variational principle
are determined by the condition that the symplectic flux 
must vanish across the spacetime boundary $B$. 
The standard Dirichlet and Neumann type boundary conditions \cite{bterms}, 
which consist of specifying 
the intrinsic metric or extrinsic curvature tensor on $B$, 
were shown to satisfy this flux condition, 
as were certain mixed type boundary conditions. 
The resulting Dirichlet boundary term gave a simple covariant derivation of
Brown and York's quasilocal energy, momentum, and angular momentum
\cite{BrownYork1,BrownYork2}. 
(These covariant results in \Refs{paperI,paperII}
should be compared with the 3+1 canonical analysis 
given in \Ref{canonical}.
Related work appears in \Ref{Nester1}.)

More importantly, this derivation showed that the Brown-York expressions
have an elegant geometrical origin
related to the spacetime mean curvature vector $\Hvec{\mu}{}$ 
\cite{ONeillI,Chen}
of the spatial boundary 2-surface $S$.
In particular, 
contraction of the time flow vector $\flowvec{\mu}{}$ 
with the dual of the mean curvature vector 
$\perpHvec{\mu}{}={*\Hvec{\mu}{}}$
in the normal space of $S$ 
yields the Brown-York energy $\varepsilon$ and outward momentum $p$ densities 
when, respectively, 
$\flowvec{\mu}{}$ is a timelike or spacelike (unit) normal vector on $S$:
$\varepsilon = -\trk_\Sigma =\flowvec{\mu}{}\perpHvec{}{\mu}$
such that $\flowvec{\mu}{}$ is orthogonal to $\Sigma$,
and $p = -\trk_B =\flowvec{\mu}{}\perpHvec{}{\mu}$
such that $\flowvec{\mu}{}$ is orthogonal to $B$,
where $\trk_\Sigma$ denotes 
the Riemannian (3-dimensional) scalar mean curvature of $S$ 
in a spacelike hypersurface spanning $\Sigma$ 
orthogonal to the timelike boundary $B$; 
likewise $\trk_B$ is the scalar mean curvature of $S$ in $B$.
Note 
$\sqrt{\varepsilon^2-p^2}
= \sqrt{-\perpHvec{\mu}{}\perpHvec{}{\mu}} =\absmcurv$
is the ``rest mass'' energy density, 
or in other words a boost-invariant mass density, 
related to the Brown-York densities 
as investigated in work of Kijowski, Lau, Epp, Liu and Yau
\cite{Kijowski,Lau,Epp,LiuYau}. 
All these densities require subtraction of a ``reference energy'' density,
typically chosen by an embedding of $S$ into a reference spacetime
interpreted as being the gravitational vacuum,
in order to yield a physically satisfactory quasilocal quantity. 
The subtraction reflects the freedom to change the Hamiltonian boundary term
by adding a function of the intrinsic metric on $S$, 
which is held fixed under the Dirichlet boundary conditions. 

Of geometrical importance, 
the mean curvature vector $\Hvec{\mu}{}$ lies in the direction 
in which the expansion of the 2-surface $S$ in spacetime is extremal.
For 2-surfaces of typical interest apart from apparent horizons,
this direction will be spacelike, 
while the direction of the dual mean curvature vector $\perpHvec{\mu}{}$ 
correspondingly will be timelike. 
In this situation the spatial covector
$\wvec{}{\mu} 
=  \absmcurv^{-2} \Hvec{}{\nu} \covSder{\mu} \perpHvec{\nu}{}$,
called the mean curvature twist, 
measures the amount of rotational-boost undergone by the pair of 
normal vectors $\Hvec{\mu}{},\perpHvec{\mu}{}$ 
when they are infinitesimally displaced in tangential directions on $S$. 
Contraction of $\wvec{}{\mu}$
with a rotational Killing vector 
$\flowvec{\mu}{}=\phi^\mu$ on $S$, provided one exists, 
directly yields the Brown-York angular momentum density:
$j_\phi= \K{\mu\nu}{} \phi^\mu {*\n}^\nu 
= \phi^\mu ( 
\absmcurv^{-1} \Hvec{}{\nu} \covSder{\mu}(\absmcurv^{-1}\perpHvec{\nu}{})
+ \covder{\mu}\chi )
=\flowvec{\mu}{}\wvec{}{\mu} +\covSder{\mu}(\chi \phi^\mu)$
(to within an irrelevant total divergence), 
where $\covSder{\mu}=\covder{\mu}|_{\coTS}$ 
denotes the spacetime derivative restricted to tangential directions on $S$;
$\K{\mu\nu}{}$ is the extrinsic curvature tensor of $\Sigma$,
$\n^\mu$ is the timelike normal to $\Sigma$,
and $\chi$ is a boost parameter on $S$ 
relating $\n^\mu$ and $\perpHvec{\mu}{}$. 
A reference subtraction can be considered for the momentum density
analogously to that for the energy density. 

The underlying geometrically defined vector 
$\Pvec{\mu}{}=\perpHvec{\mu}{}+\wvec{\mu}{}$ here,
arising from the covariant gravitational Hamiltonian, 
is called \cite{paperI} the Dirichlet symplectic vector of $S$. 
It depends only on the extrinsic geometry of $S$ in spacetime, 
and its coordinate components can be computed from 
the gravitational initial data on any spacelike hypersurface that spans $S$. 
Properties and examples of this vector for spacelike 2-spheres 
in spacetimes of physical interest 
(isometry 2-spheres in spherically symmetric spacetimes 
and homogeneous isotropic spacetimes, 
constant-radius spheres in axisymmetric spacetimes, 
large spheres in asymptotically flat spacetimes,
and spherical surfaces in flat-spacetime hyperplanes and light cones)
have been studied in \Ref{paperII}. 

The main idea of the present paper is to study mathematical aspects of 
the dual mean curvature vector $\perpHvec{\mu}{}$ 
and the mean curvature twist vector $\wvec{\mu}{}$ 
as gravitational Hamiltonian flows of 2-surfaces in spacetime. 
In \secref{geometry},
geometrical properties of these vectors are developed,
particularly connected with apparent horizons and Killing vectors, 
stemming from the fact that the expansion of 
any spacelike 2-surface $S$ in spacetime
is found to vanish in the direction of $\perpHvec{\mu}{}$. 
A covariant analysis of Hamiltonian boundary terms 
for the mean curvature flows 
$\absperpmcurv^{n-1} \perpHvec{\mu}{}$ and $|\mcurv|^{n}\wvec{\mu}{}$,
as well as their natural generalization $\absP^{n-1} \Pvec{\mu}{}$, 
is carried out for spacelike 2-surfaces with timelike $\perpHvec{\mu}{}$
in \secref{analysis}.
The boundary conditions associated with these flows are also discussed,
along with choices for ``reference'' (vacuum) subtraction terms 
in the boundary Hamiltonian
depending on the gravitational initial data at $S$ 
held fixed by these boundary conditions. 
A generalization of the analysis and results is also stated
for spacelike 2-surfaces with null $\perpHvec{\mu}{}$,
thereby encompassing the interesting case of apparent horizons. 
Geometrical definitions of 
mean curvature quasilocal mass and angular momentum
given by the resulting Hamiltonians are then considered 
in \secref{quasilocal}.
This yields a one-parameter ($n$) family of quasilocal masses 
that is shown to include both 
the Hawking mass ($n=1$)
and the Brown-York ``rest mass'' energy ($n=0$). 
In \secref{properties}, 
firstly, 
the large sphere limit of these definitions is shown to yield
the ADM mass and angular momentum in asymptotically flat spacetimes,
and a lower bound in terms of irreducible mass is established 
at apparent horizons. 
Secondly, 
a positivity result is proved for the quasilocal masses,
motivated by consideration of spacelike mean curvature flow of
2-surfaces in spacetime. 
This flow is used to reduce the positivity statement for the case $n=0$
to a recent theorem on Riemannian mean curvature \cite{ShiTam},
and then a (negative) lower bound is derived in the cases $n\geq 1$. 

An index-free notation \cite{ONeillI} is mainly used hereafter.

\section{ Geometry of spacelike 2-surfaces in spacetime }
\label{geometry}

First, recall \cite{ONeillI,Chen} that 
the geometry of spacelike 2-surfaces $S$ 
embedded in a 4-dimensional spacetime $(M,g)$ 
is characterized by the intrinsic metric, $\sigma$, 
(first fundamental form) of $S$, 
and the extrinsic shape tensor, $\Pi$, 
associated with the extrinsic curvature tensor, $k$, 
(second fundamental form)
for any normal vector $\e{\perp}{}$ of $S$ via
$k= -g(\e{\perp}{},\Pi)$.
The metric $\sigma=g|_{\coTS}$ is given by the pullback of $g$ to $S$,
\ie/ $\sigma(v,w)=v\cdot w$
for all tangential vectors $v,w$, 
while the shape tensor is defined by 
$\Pi(v,w)= (\covder{v} w)|_{\TperpS} = (\covder{w} v)|_{\TperpS}$
where $\covder{}$ denotes the covariant derivative for $(M,g)$.
Throughout, the inner product of vectors with respect to the spacetime metric
will be denoted by $v\cdot w=g(v,w)$.
The inner product of a vector with itself will be written
$v^2=v\cdot v=g(v,v)$,
and $|v|=\sqrt{|v\cdot v|}$ will denote the absolute norm of a vector $v$.

For any spacelike 2-surface $S$ in spacetime $(M,g)$, 
at each point on $S$ there is geometrically preferred normal vector, 
defined by the mean curvature vector of $(S,\sigma)$, 
\EQ
\mcurv = \trk(\e{1}{}) \e{1}{} - \trk(\e{0}{}) \e{0}{}
\doneEQ
where $\trk(\e{0}{})$ and $\trk(\e{1}{})$ 
are the extrinsic scalar curvatures of $S$
(trace of the second fundamental forms) 
in the directions of any orthonormal frame 
$\{\e{0}{},\e{1}{}\}$ of $\TperpS$. 
The vector $\mcurv$ is independent of choice of normal frame for $S$,
\ie/ it is invariant under boosts and reflections of $\e{0}{},\e{1}{}$, 
and hence is well-defined given just the 2-surface 
and its extrinsic geometry in spacetime. 
The absolute norm of $\mcurv$ is the scalar mean curvature of $S$
\EQ
\absmcurv = \sqrt{|\trk(\e{1}{})^2-\trk(\e{0}{})^2|}
\doneEQ
which covariantly generalizes the definition of mean curvature 
in Euclidean surface theory.
More specifically, 
in Minkowski space if $S$ lies in a spacelike hyperplane then 
$\mcurv$ is its Euclidean mean curvature vector orthogonal to $S$
and, in particular, 
$\absmcurv$ is the Euclidean extrinsic mean curvature of $S$. 
(Note, in $n$-dimensional Riemannian geometry \cite{Chen}
the mean curvature vector of a 2-surface is commonly defined by 
$\frac{1}{n}\sum_{i=1}^{n-2} \trk(\e{i}{}) \e{i}{}$
in terms of a normal frame $\e{i}{}$. 
Apart from an obvious sign change due to the Lorentzian signature, 
this differs by a factor $\frac{1}{2}$ 
compared with the definition of $\mcurv$ used here.)
Hereafter it will be assumed that $S$ is closed and $M$ is orientable. 
The sign conventions used throughout will be that,
with respect to a smooth spacelike hypersurface $\Sigma$ spanning $S$
for which $\e{1}{}$ is tangent and $\e{0}{}$ is normal, 
$\e{1}{}$ is chosen outward pointing, $\e{0}{}$ is chosen future pointing, 
and $\trk(\e{1}{})$ is chosen positive 
if the interior of $S$ is a convex domain 
so then $\absmcurv=\trk(\e{1}{})>0$ for such a 2-surface $S$. 
This choice of frame for $\TperpS$ will be called an oriented normal frame. 

Now introduce a unique normal vector orthogonal to $\mcurv$, 
given by 
\EQ
\perpmcurv = \trk(\e{1}{}) \e{0}{} - \trk(\e{0}{}) \e{1}{}
\doneEQ
which will be called 
the {\it dual mean curvature vector} of $(S,\sigma)$,
as it is dual to $\mcurv$ in $\TperpS$, 
\EQ
{*\mcurv}=\perpmcurv ,\quad {*\perpmcurv}=\mcurv .
\doneEQ
The vector $\perpmcurv$ has the same absolute norm as $\mcurv$, 
and for $S$ lying in a spacelike hyperplane in Minkowski space,
its direction is that of a timelike normal to the hyperplane. 

Like $\mcurv$, the normal vector $\perpmcurv$ does not depend on 
the choice of a normal frame for $\TperpS$
and so is well-defined given just the 2-surface and its extrinsic geometry. 
Associated to this pair of geometrical normal vectors $\mcurv,\perpmcurv$
is a tangential covector $\twist(\mcurv)$ given by 
\EQ\label{scaledtwist}
v\hook \twist(\mcurv) 
= \mcurv\cdot \covder{v}\perpmcurv
= -\perpmcurv\cdot \covder{v}\mcurv
\doneEQ
for all tangent vectors $v$ on $S$.
Note $\twist(\mcurv)$ and its dual $*\twist(\mcurv)$
are determined just by the intrinsic and extrinsic geometry of $S$
in spacetime and hence define geometrically preferred
tangential covectors on $S$, 
respectively called the {\it twist} and {\it dual twist}
of the mean curvature vectors, 
with $*\twist(\mcurv)$ being orthogonal to $\twist(\mcurv)$. 

An important geometrical property of $\perpmcurv$ is that 
the intrinsic expansion of $S$ in the direction of $\perpmcurv$ 
in any spacetime $(M,g)$ is zero, 
\EQ\label{zeroexp}
(\Lie{\perpmcurv} \vol{}{}(S) )|_{\coTS} =0
\doneEQ
where $\vol{}{}(S)$ is the area element 2-form of $S$ 
normalized by the metric $\sigma$ on $S$. 
This property follows from the fact that 
the scalar extrinsic curvatures of $S$ 
in the directions $\mcurv$ and $\perpmcurv$
obey 
\EQ\label{mcurvexp}
\trk(\perpmcurv)=0 ,\quad
\trk(\mcurv) = \mcurv^2 = -\perpmcurv^2 ,
\doneEQ
where $\trk(v) \equiv \frac{1}{2} g(\sigma,\Lie{v}\sigma)$
defines the scalar extrinsic curvature with respect to any normal vector
$v$ in $\TperpS$. 
Moreover, the norm of $\perpmcurv$ is related to the property of 
trapping of null geodesics through $S$. 
Consider outward and inward null vectors $\outnull,\innull$ at $S$,
with the normalization $\outnull\cdot \innull=-1$. 
In terms of the null frame $\{\outnull,\innull\}$, 
the vectors $\mcurv$ and $\perpmcurv$
are expressed by 
\EQs
\mcurv = -\trk(\outnull) \innull -\trk(\innull) \outnull ,\quad
\perpmcurv = \trk(\outnull) \innull -\trk(\innull) \outnull , 
\doneEQs
with norms 
\EQ
\perpmcurv^2 = 2 \trk(\outnull) \trk(\innull) = -\mcurv^2 .
\doneEQ
Recall \cite{Wald-book}, 
$\trk(\outnull)$ and $\trk(\innull)$ 
measure the convergence of a congruence of 
future directed (outward and inward) null geodesics at $S$, 
and the condition for these geodesics to be trapped at $S$ is that 
the expansion of $S$ in both the outward and inward null directions is
negative, $\trk(\outnull)<0$ and $\trk(\innull)<0$ 
everywhere on $S$. 
Consequently, the situation when $\perpmcurv^2 =0$ holds everywhere on $S$
characterizes $S$ as being a marginally-outward or -inward trapped surface
in spacetime. 

Let $\Sigma$ be a smooth spacelike hypersurface spanning $S$. 
The condition on $S$ that $\mcurv=\perpmcurv$ in $\TM|_S$,
namely $\trk(\outnull)=0$ at every point on $S$, 
describes a marginally-outward trapped surface \cite{HawkingEllis},
sometimes also referred to as an apparent horizon on $\Sigma$. 
Thus for such a surface $S\subset \Sigma$, 
$\perpmcurv$ and $\mcurv$ degenerate into a single null vector. 
If instead $\mcurv^2< 0$ holds at every point on $S$,
whereby $\mcurv$ is timelike while $\perpmcurv$ is spacelike
in $\TM|_S$, 
so $\trk(\outnull)$ and $\trk(\innull)$ 
have the same (nonzero) sign on $S$
then $S\subset \Sigma$ is said to be a strictly (or closed) trapped surface 
\cite{HawkingEllis}. 
Future and past (strictly) trapped surfaces $S$ are distinguished
according to whether $\trk(\innull)$ is positive or negative on $S$. 
In the context of formation of black holes 
in General Relativity \cite{HawkingEllis,Wald-book}, 
trapped surfaces $S\subset \Sigma$ obey 
$\trk(\outnull)\leq 0$ and $\trk(\innull)<0$ everywhere on $S$,
corresponding to the condition that in $\TM|_S$
the null mean curvature vector $\perpmcurv+\mcurv$ is future pointing
and the mean curvature vector $\mcurv$ is timelike or null
\cite{ONeillI}. 

Suppose that every hypersurface $\Sigma$ spanning $S$ 
contains no trapped surfaces in its interior
(hence $S$ does not surround any black hole horizons in spacetime). 
Then, in physically reasonable situations, 
$\mcurv$ can be expected to be spacelike or null in $\TM|_S$.
If $\perpmcurv$ is future pointing in $\TM|_S$, 
then given any hypersurface $\Sigma$ chosen orthogonal to $\perpmcurv$, 
the direction of $\mcurv$ in $\Sigma$ will be outward or inward
at a given point according to whether 
$S\subset \Sigma$ is convex or concave at that point. 
Thus, surfaces $S$ obeying $\mcurv^2 >0$ at every point
are necessarily a convex boundary for any spanning spacelike hypersurface. 

Definition~2.1: 
A 2-surface $S$ will be called {\it regular} 
if it is closed, smooth, and spacelike, 
and if $\mcurv^2 \geq 0$ (or equivalently $\perpmcurv^2 \leq 0$) 
holds at every point on the surface.
If $\mcurv^2>0$ at every point on $S$, 
it will be called {\it convex-regular},
and when such a 2-surface is a convex boundary for a spanning hypersurface, 
it will be called a regular {\it convex-boundary} surface. 

Regularity is a relatively weak requirement on 2-surfaces,
in particular, it does not restrict the convexity of $S$,
since points at which $\absmcurv=0$ encompass any indentations 
(change of convexity) on $S$;
also it includes apparent horizons,
as well as the related notions of 
isolated horizons and dynamical horizons in spacetime, 
whose spatial cross-sections $S$ are required (among other properties)
to obey $\absmcurv=0$ at every point,
as introduced recently by Ashtekar and coworkers
\cite{horizons}. 

The rest of the extrinsic curvature of $S$ is described by 
the shears of (inward/outward) null geodesic congruences at $S$,
namely the trace-free part of the second fundamental forms 
in the null normal directions. 
It will be convenient to regard the second fundamental form of $S$
in a given normal direction $\e{\perp}{}$ 
as a 1-form $\k{}{}(\e{\perp}{})$ in $\coTS$ with values in $\TS$
defined by
\EQ\label{shear}
w\cdot(v\hook \k{}{}(\e{\perp}{}) ) = -\e{\perp}{} \cdot \covder{v}w
= -\e{\perp}{} \cdot \Pi(v,w)
\doneEQ
for all tangent vectors $v,w$ in $\TS$. 
Note there is a decomposition of $\k{}{}(\e{\perp}{})$ 
into trace-free and trace parts with respect to the metric 
$\metric{}{}$ on $S$, both of which are symmetric tensors, 
given by 
\EQ
\tr\k{}{}(\e{\perp}{}) 
\equiv \frac{1}{2} \trk(\e{\perp}{}) \metric{}{} ,\quad
\k{}{}(\e{\perp}{}) - \tr\k{}{}(\e{\perp}{}) 
\equiv \shear{}{}(\e{\perp}{}) .
\doneEQ
In the mean curvature directions for a regular 2-surface $S$,
the second fundamental forms satisfy 
\EQ
\tr\k{}{}(\perpmcurv) = 0 ,\quad
\tr\k{}{}(\mcurv) = \frac{1}{2} \absmcurv^2 \metric{}{} .
\doneEQ
Thus the extrinsic curvature is completely determined by 
the scalar mean curvature $\absmcurv$
and the shears 
$\shear{}{}(\perpmcurv)$, $\shear{}{}(\mcurv)$. 
Recall,  
a 2-surface is umbilic \cite{ONeillI,Chen}
with respect to a normal direction $\e{\perp}{}$ 
at a point if 
$(\covder{v} w)|_{\TperpS} \propto {v\cdot w}\ \e{\perp}{}$. 
It follows from equation \eqref{shear} that 
$\k{}{}(\e{\perp}{})$ is pure trace at such a point, 
\ie/ $S$ is shear free in an umbilic direction. 
This leads to a characterization of the mean curvature shears
as measuring the deviation of $S$ from being umbilic 
in the mean curvature normal directions at any point on $S$. 

Proposition~2.2:
A regular 2-surface $S$ is shear free in all normal directions in spacetime
iff it is umbilic with respect to the spacelike mean curvature direction 
$\mcurv$ at every point on the surface. 

Another extrinsic aspect of the geometry of $S$ in $(M,g)$ is
the curvature of its normal space $\TperpS$ as described by 
the sectional normal curvature tensor \cite{Chen}.
This normal curvature turns out to be directly determined by
the twist $\twist(\mcurv)$ of the mean curvature vectors $\mcurv,\perpmcurv$,
as follows. 

Firstly, 
the normal curvature tensor is introduced 
in terms of the normal covariant derivative $\perpSder{}$ on $S$ by
\EQ
\norScurv{}{}(v) = [\perpSder{},\perpSder{}](v)
\doneEQ
where
\EQ
\perpSder{w} v = (\covder{w} v)|_{\TperpS}
\doneEQ
for all normal and tangential vector fields $v$ and $w$ on $S$.
Note $\norScurv{}{}(\cdot)$ is a 
$\coTperpS \otimes \TperpS$-valued 2-form on $S$. 
Geometrically, it describes the curvature of the normal bundle 
$(\TperpS,\perpmetric{}{})$ of $S$. 
In comparison, 
the intrinsic curvature of $S$ is obtained from 
the tangential covariant derivative $\Sder{}$ on $S$
\EQ
\Sder{w} v = (\covder{w} v)|_{\TS}
\doneEQ
for all tangential vector fields $v$ and $w$ on $S$,
giving the intrinsic curvature tensor
\EQ
\curvS{}{}(v) = [\Sder{},\Sder{}](v)
\doneEQ
defined as a $\coTS \otimes \TS$-valued 2-form $\curvS{}{}(\cdot)$ on $S$. 
Note, because $S$ is two dimensional,
the curvature tensors have the tensor product form 
\EQ
\curvS{}{}(v) = \curvS{}{} \vol{}{}(S)\otimes{*v} ,\quad
\norScurv{}{}(v) = \norScurv{}{} \vol{}{}(S)\otimes{*v} ,
\doneEQ
for $v$ in $\TS$ and $\TperpS$, respectively, 
with $\curvS{}{}$ and $\norScurv{}{}$ denoting scalar curvatures;
in particular, $\curvS{}{}$ is the Gaussian curvature of $S$. 
Here the dual operation $*$ on $\TS$ as well as on $\TperpS$ 
is defined in the standard way 
in terms of the linear maps associated with the 2-forms 
$\vol{}{}(S)$ and $\perpvol{}{}(S)=\e{1}{}\wedge\e{0}{}$ 
related such that 
$\vol{}{}(S) \wedge \perpvol{}{}(S) =\vol{}{}$ is 
the spacetime volume 4-form normalized with respect to the metric $g$.

Now consider 
the fundamental normal form $\twist(e)$ of $S$, 
defined by \cite{Chen}
\EQ
v \hook \twist(e) = \e{1}{} \cdot \covder{v} \e{0}{} 
\doneEQ
for all tangent vector fields $v$ on $S$. 
Note $\twist(e)$ is a 1-form in $\coTS$
depending on a choice of orthonormal frame $\{\e{0}{},\e{1}{}\}$ 
for $\TperpS$,
with the resulting transformation property 
\EQ
\twist(e) \rightarrow \twist(e) + \Sder{}\chi
\doneEQ
under boosts of the normal frame in $\TperpS$
\EQ\label{boost}
\e{0}{} \rightarrow \cosh\chi\ \e{0}{} +\sinh\chi\ \e{1}{} ,\quad
\e{1}{} \rightarrow \cosh\chi\ \e{1}{} +\sinh\chi\ \e{0}{}
\doneEQ
where $\chi$ is the boost parameter given by a smooth function on $S$. 
Such transformations represent
a $SO(1,1)$ gauge group acting on $\TperpS$,
with $\twist(e)$ transforming as a $SO(1,1)$ connection 1-form. 
The gauge invariant curl of $\twist(e)$, 
representing the curvature 2-form of this gauge group,
is found to coincide with 
the normal curvature 2-form
\EQ
\norScurv{}{}\vol{}{}(S) = \d{}_S\twist(e)
\doneEQ
where $\d{}_S$ denotes the exterior derivative on $S$. 

The mean curvature vectors $\mcurv,\perpmcurv$ provide a geometrically
preferred normal frame for $\TperpS$
and hence also a corresponding preferred normal connection 
related to the twist $\twist(\mcurv)$,
for any spacelike 2-surface $S$.

To begin, let $S$ be a convex-regular 2-surface,
so then $\perpmcurv$ provides a geometrically preferred 
timelike vector field in $\TperpS$. 
This yields the preferred normal frame 
\EQ
\unitperpmcurv \equiv \absmcurv^{-1} \perpmcurv, \quad
\unitmcurv \equiv \absmcurv^{-1} \mcurv ,\quad
-\unitperpmcurv^2 = \unitmcurv^2 = 1 ,\quad
\unitperpmcurv\cdot \unitmcurv = 0 ,
\doneEQ
which will be called the {\it mean curvature frame} of $\TperpS$,
along with a corresponding preferred null frame 
\EQ
\unitmcurv_{\pm} 
\equiv \frac{1}{\sqrt{2}}( \unitperpmcurv \pm \unitmcurv ) ,\quad
\unitmcurv_{\pm}^2=0 ,\quad
\unitmcurv_{+}\cdot\unitmcurv_{-} =-1 . 
\doneEQ
The mean curvature frame gives a way to gauge fix $\twist(e)$,
yielding a geometrically preferred normal connection $\twist$ given by 
\EQ
v \hook \twist = \unitmcurv \cdot \covSder{v} \unitperpmcurv
\doneEQ
which is related to the twist of the mean curvature vectors
\EQ
\twist = \absmcurv^{-2}\twist(\mcurv)
\doneEQ
and hence will be called the {\it mean curvature twist covector} 
of $(S,\sigma)$. 
Geometrically, 
$\twist$ is a well-defined tangential covector field on $S$
and measures the magnitude of 
the $SO(1,1)$ rotation of the mean curvature frame 
under infinitesimal displacement on $S$.
It determines the normal curvature of $S$ by 
\EQ
\norScurv{}{}\vol{}{}(S) =\d{}_S\twist 
= \absmcurv^{-2}( \d{}_S\twist(\mcurv) -
2\d{}_S\ln\absmcurv\wedge\twist(\mcurv) ) .
\doneEQ
If $S$ is Lie dragged separately 
along the directions $\mcurv,\perpmcurv$ in $(M,g)$ 
to produce a pair of hypersurfaces that intersect orthogonally at $S$,
then $\twist$ is directly related to the tangential part of 
the mean curvature frame commutator
\EQ
v \hook \twist = \frac{1}{2} v \cdot [\unitmcurv,\unitperpmcurv]|_S
\doneEQ
for all vectors $v$ in $\TS$. 
In this situation, 
$\twist$ is also related to the extrinsic expansion of $S$ 
along the timelike direction $\perpmcurv$, 
\EQ
\Lie{\perpmcurv} \vol{}{}(S) = 2 {*\twist} \wedge \mcurv
\doneEQ
where ${*\twist}$ denotes the dual of $\twist$ on $S$
and $\mcurv$ is viewed here as a 1-form 
through the identification of $\TperpS$ and $\coTperpS$
provided by the metric $\perpmetric{}{} = g|_{\coTperpS}$ 
on the normal space of $S$. 

These results can be extended to spacelike 2-surfaces $S$ that 
are merely regular, 
thus encompassing apparent horizons and marginally-outward trapped surfaces
characterized respectively by $\absmcurv=0$ on $S$, and 
$\mcurv=\perpmcurv$ on $S$. 
For a regular 2-surface $S$, $\perpmcurv$ provides a 
geometrically preferred non-spacelike vector field in $\TperpS$.
Associated to $\perpmcurv$ is a null vector field given by 
\EQ
\mcurv_+ \equiv \frac{1}{2}(\mcurv +\perpmcurv), \quad
\mcurv_+^2=0 ,
\doneEQ
so that in the case of an outward trapped surface, 
$\mcurv_+ = \mcurv =\perpmcurv$. 
In terms of any orthonormal frame $\{\e{0}{},\e{1}{}\}$, 
note
\EQ\label{outnull+}
\mcurv_+ = \frac{1}{2}(\trk(\e{1}{}) - \trk(\e{0}{}))( \e{0}{}+\e{1}{} )
\doneEQ
is invariant under boosts \eqref{boost} since
\EQ
\e{0}{} \pm \e{1}{} \rightarrow \exp(\pm\chi) ( \e{0}{} \pm \e{1}{} ) .
\doneEQ
Now introduce a transverse null vector field 
\EQ\label{outnull-}
\mcurv_- \equiv (\trk(\e{1}{}) - \trk(\e{0}{}))^{-1}( \e{0}{}-\e{1}{} )
\doneEQ
likewise invariant under boosts,
and obeying 
\EQ
\mcurv_-^2=0 ,\quad
\mcurv_-\cdot\mcurv_+ =-1 .
\doneEQ
Note the scalar extrinsic curvatures of $S$ in these null directions
are given by
\EQ
\trk{}{}(\mcurv_+) = \frac{1}{2} \mcurv^2 ,\quad
\trk{}{}(\mcurv_-) = -1 .
\doneEQ
In the case of an inward trapped surface where $\mcurv =-\perpmcurv$ on $S$,
the expressions for $\mcurv_\pm$ must be changed to 
\EQ\label{innull-}
\mcurv_+ \equiv \frac{1}{2}(\perpmcurv -\mcurv)
= \frac{1}{2}(\trk(\e{1}{}) + \trk(\e{0}{}))( \e{0}{}-\e{1}{} )
\doneEQ
and 
\EQ\label{innull+}
\mcurv_- \equiv (\trk(\e{1}{}) + \trk(\e{0}{}))^{-1}( \e{0}{}+\e{1}{} ) ,
\doneEQ
while accordingly
$\trk{}{}(\mcurv_+) = -\frac{1}{2} \mcurv^2$
and $\trk{}{}(\mcurv_-) = 1$. 
Note, in both the inward- and outward-trapped cases,
the null vectors $\mcurv_\pm$ have the orientation
$\perpvol{}{}(S) =\mcurv_+\wedge\mcurv_-$. 

The pair of normal vectors $\{\mcurv_\pm\}$ thereby provides
a geometrically preferred null frame for $\TperpS$,
which is well-defined just given $S$ and its extrinsic geometry in $(M,g)$
such that $\absmcurv \ge 0$.
This {\it mean curvature null frame} determines a corresponding
preferred (gauge fixed) connection $\twist^\pm$, 
such that for all vectors $v$ in $\TS$,
\EQ\label{nullmcurvtwistout}
v \hook \twist^+
= \mcurv_+ \cdot \covSder{v} \mcurv_-
\doneEQ
in the outward-trapped case \eqrefs{outnull+}{outnull-},
and similarly
\EQ\label{nullmcurvtwistin}
v \hook \twist^-
= \mcurv_- \cdot \covSder{v} \mcurv_+
\doneEQ
in the inward-trapped case \eqrefs{innull-}{innull+}.
The 1-forms $\twist^\pm$ will be called the {\it null mean curvature twist}.
They have geometrical properties similar to those of $\twist$.
In particular, firstly, 
the normal curvature of $S$ is given by 
\EQ
\pm \norScurv{}{}\vol{}{}(S)= \d{}_S\twist^\pm .
\doneEQ
Secondly, 
If $S$ is Lie dragged separately 
along the null directions $\mcurv_\pm$ in $(M,g)$ 
to produce a pair of null hypersurfaces that intersect orthogonally at $S$,
then the null mean curvature frame commutator is related to $\twist^\pm$ via
\EQ
v \cdot [\mcurv_+,\mcurv_-]|_S
= \pm 2 v \hook \twist^\pm + \covSder{v}(f_- -f_+)|_S
\doneEQ
for all vectors $v$ in $\TS$;
here $f_\pm$ are functions coming from the hypersurface orthogonality
conditions \cite{Wald-book} $\d{}\mcurv_\pm =\d{}f_\pm \wedge \mcurv_\pm$. 
Thirdly, in this situation, 
$\twist^\pm$ is also related to the expansion of $S$ 
along the null direction $\mcurv_+$ in spacetime, 
\EQ
\Lie{\mcurv_+} \vol{}{}(S) = 
\frac{1}{2}\absmcurv^2 \vol{}{}(S) 
+ 2 {*[\mcurv_+,\mcurv_-]|_S} \wedge \mcurv_+
\doneEQ
where $*$ denotes the dual of the commutator viewed as a 1-form 
through the identification of $\TperpS$ and $\coTperpS$
provided by the metric $\perpmetric{}{} = g|_{\coTperpS}$ 
on the normal space of $S$. 

For any convex-regular 2-surface $S$, 
a comparison of $\twist$ and $\twist^\pm$ yields the relation 
\EQ
\twist_+ =-\twist_-=\twist -\Sder{}(\smallfrac{1}{2}\ln\absmcurv)
\doneEQ
reflecting the different gauge choices for the previous normal frames
in $\TperpS$. 

Definition~2.2: 
A 2-surface $S$ is said to have constant mean curvature (CMC) 
if $\mcurv^2=\const$ on $S$.

It immediately follows that for a positive CMC 2-surface $S$,
$\twist^\pm =\pm\twist$, 
so there is then a unique geometrically preferred connection 1-form
for the normal curvature of $S$. 
If $S$ is instead a zero CMC 2-surface, then $\twist$ becomes singular
(since $\absmcurv=0$) while $\twist^\pm$ represents its non-singular part. 

Proposition~2.3:
The mean curvature vectors $\mcurv$, $\perpmcurv$, 
and null frame $\mcurv_\pm$, 
along with the null twist $\twist^\pm$ covector, 
are geometrically well-defined for any regular spacelike 2-surface $S$,
including apparent horizons, 
and depend only on the intrinsic and extrinsic geometry of $S$ in $(M,g)$.
For convex-regular 2-surfaces $S$, 
there is geometrically well-defined normal frame $\{\perpmcurv,\mcurv\}$
and corresponding twist covector $\twist$. 

Finally, 
there is an interesting relationship between mean curvature vectors
and Killing vectors. 
Suppose $\kv$ is a Killing vector of the spacetime metric $g$,
$\Lie{\kv} g=0$. 
Projection of the Killing equation into the tangent space of 
any spacelike 2-surface $S$ yields $\Lie{\kv} \sigma|_\coTS =0$
and hence $g(\sigma,\Lie{\kv} \sigma) =0$,
implying $S$ has non-vanishing intrinsic expansion 
in the Killing vector direction in spacetime,
\EQ
\trk(\kv)=0 .
\doneEQ
A comparison of this property with the main geometric property \eqref{mcurvexp}
of mean curvature vectors leads to the following result.

Proposition~2.4:
Suppose $\kv$ is a Killing vector orthogonal to $S$ in $(M,g)$.
Then $\kv$ at $S$ is proportional to the dual mean curvature vector of $S$,
\EQ
\kv|_S = \alpha\perpmcurv
\doneEQ
for some smooth non-vanishing function $\alpha$ on $S$.

Such a Killing vector is obviously either timelike or null at $S$. 
(Hereafter, without loss of generality, $\kv|_S$ and $\perpmcurv$
will be assumed to share the same orientation in the normal space of $S$,
so $\alpha>0$ on $S$.)
In the timelike case, $\kv^2|_S <0$, 
the mean curvature vector $\mcurv$ is necessarily spacelike
since it is orthogonal to $\kv$, 
so the dual vector $*\kv$ is thereby proportional to $\mcurv$,
\EQ
*\kv|_S = \alpha\mcurv . 
\doneEQ
In particular, note the norms of these vectors are related by
$|\kv|_S=\alpha\absmcurv$. 
It now follows that the twist $\twist(\mcurv)$ of the mean curvature vectors
is similarly related to the twist of the normal frame provided by
the Killing vector and its dual $\{\kv,*\kv\}$:
\EQ
v\hook\twist(\kv) = ({*\kv}\cdot\covSder{v}\kv)|_S
= \alpha^2\mcurv\cdot\covSder{v}\perpmcurv
\doneEQ
and thus
\EQ
\twist = |\kv|^{-2}\twist(\kv) .
\doneEQ
A similar geometric result holds in the case of 
a Killing vector that is null at $S$, $\kv^2|_S =0$.
Since both $\perpmcurv$ and $\mcurv$ are then necessarily null vectors,
with $\mcurv_\pm = \perpmcurv = \pm \mcurv$
(in the outward/inward trapped cases respectively),
the twist $\twist(\kv)$ of a null frame aligned with the null vector 
$\kv|_S =\alpha\mcurv_\pm$
differs from the null mean curvature twist just by a gradient
\EQ
\twist^\pm =\twist(\kv) +\covSder{}(\ln\alpha)
\doneEQ
corresponding to a boost relating the null vectors $\kv$ and $\mcurv_\pm$. 

Further discussion of these various mean curvature 1-forms and vectors
in the convex-regular case,
as well as their calculation in spacetimes of interest, 
can be found in \Ref{paperII}. 
Hereafter, 
$\mcurv,\perpmcurv,\twist,\perptwist$
will be all regarded as vectors or covectors,
depending on the geometrical context,
through the identification of $\TS$ with $\coTS$,
and $\TperpS$ with $\coTperpS$, 
using the metrics $\metric{}{}$ and $\perpmetric{}{}$. 


\section{ Hamiltonian analysis with spatial boundary\\ 
conditions related to mean curvature flow }
\label{analysis}

The gravitational ADM Hamiltonian provides a variational principle whose
stationary points yield the Einstein field equations 
in a 3+1 dynamical form \cite{ADM,Wald-book}. 
This formulation requires specifying 
a time-flow vector field $\flow$
on spacetime together with 
choosing a foliation given by spacelike slices $\Sigma$. 
Suppose a mathematically defined boundary is introduced, 
enclosing a spatially compact region of spacetime 
with a timelike boundary hypersurface $B$. 
Such boundaries arise 
in numerical relativity where the Cauchy data must be cutoff
on noncompact slices, 
and in astrophysical contexts that involve separating a spatial region
into a finite-size system and its exterior, 
and also in gluing of local solutions for the Cauchy problem.

In the approach taken in \Refs{BrownYork1,BrownYork2,Kijowski,paperI,paperII}
on the Hamiltonian structure of the Einstein field equations, 
boundaries were restricted to be 
a fixed (non-dynamical) timelike hypersurface $B$ in spacetime
and the time-flow was chosen as some fixed vector field $\flow$ 
in the hypersurface. 
Let $S=\Sigma \cap B$ denote the spacelike 2-surfaces 
foliating the boundary $B$, which will be viewed as 
defining a spatial boundary $\partial\Sigma=S$ on each slice $\Sigma$. 
Let $\n$ denote the future pointing unit normal to $\Sigma$. 
(It is not assumed that $\Sigma$ is orthogonal to $B$). 
Then, the ADM Hamiltonian 
\EQ\label{ADMH}
\H{\rm ADM}(\xi) = \frac{1}{8\pi}
\int_\Sigma \Hdens \d{\Sigma} ,\quad
\Hdens = 
\ricci(\flow,\n) - \frac{1}{2} \g{}{}(\flow,\n) \scurv
\doneEQ
provides a variational principle that yields 
the 3+1 form of the Einstein field equations 
under spatially compact support variations of the gravitational field 
variables on the interior of $\Sigma$. 
However, 
when variations are considered that have nonvanishing support 
at the boundary $\partial\Sigma$, 
the variational derivatives of the ADM Hamiltonian 
will no longer be well-defined due to a boundary term that arises
through the presence of the boundary surface. 
Spatial boundary conditions on the gravitational field variables 
at this surface $\partial\Sigma$ must then be sought that allow 
modifying the ADM Hamiltonian by addition of a surface integral 
so that, under variations with support on $\partial\Sigma$, 
a well-defined Hamiltonian variational principle is obtained
for the 3+1 form of the Einstein field equations. 

An analysis of allowed boundary conditions and compensating boundary terms
is made in \Refs{paperI,paperII}
with a covariant symplectic approach that uses
the Noether charge method developed by Wald and coworkers
\cite{noethercharge1,noethercharge2,noethercharge3}. 
The results are summarized as follows. 
Let the spacetime metric be decomposed as
$g=\sigma-\e{0}{}\otimes\e{0}{}+ \e{1}{}\otimes\e{1}{}$
where $\{\e{0}{},\e{1}{}\}$ is a oriented normal frame of $S$ adapted to $B$. 
Boundary conditions for the existence of a Hamiltonian are determined by
the condition that the symplectic flux across $B$ must vanish. 
This condition is satisfied by Dirichlet and Neumann boundary conditions 
consisting of fixing 
either the induced metric $q=\sigma -\e{0}{}\otimes\e{0}{}$ on $B$ 
or the extrinsic curvature tensor $p=\frac{1}{2} \Lie{e_1}q$ on $B$
(\ie/ the first or second fundamental forms of $B$ in $(M,g)$). 
As well, some other mixed types of boundary conditions are allowed. 

Theorem~3.1:
Fix a vector field $\flow$ tangent to $B$ and independent of $g$,
and consider the Dirichlet boundary condition on $g$ at $\partial\Sigma$. 
Let $\xi$ be extended smoothly from $\partial\Sigma$ into $\Sigma$. 
Then a Hamiltonian for the 3+1 form of the Einstein field equations on $\Sigma$
is given by 
\EQ
\H{}(\flow) = 
\H{\rm ADM}(\xi) 
+ \frac{1}{8\pi}
\oint_{\partial\Sigma} \flow\cdot( \perpmcurv +\twist_B +\refHdens{}) \d{S}
\doneEQ
where $\refHdens{}$ depends only on the boundary hypersurface metric $q$
at $\partial\Sigma$
and corresponds to the freedom to add 
an arbitrary function of the boundary data $\sigma=q|_{\Tsurf}$
to the Hamiltonian;
here $\perpmcurv$ is the dual mean curvature vector of $\partial\Sigma$
in $\Tperpsurf$
and $\twist_B$ is the twist vector in $\Tsurf$
defined from an orthonormal frame adapted to $B$.
The variational derivatives $\delta \H{}(\flow)/\delta g$ 
are well-defined and give as stationary points of $\H{}(\flow)$
the 3+1 form of the Einstein field equations
with respect to $\Sigma$.

This theorem is proved in \cite{paperI}.
For a comparison with related results in the literature, 
it is useful to summarize the variational form of 
the gravitational field equations provided by the Hamiltonian $\H{}(\flow)$.

On $\Sigma$, the gravitational field variables (\ie/ Cauchy data) consist of 
the spatial metric tensor $\h{}{}=g-\n\otimes\n$ 
and the extrinsic curvature tensor $\K{}{}=\frac{1}{2}\Lie{\n}\h{}{}$. 
It is convenient to decompose the time-flow vector into 
normal and tangential parts $\flow = \lapse \n +\shift{}{}$ on $\Sigma$.
In coordinates $\{\x{i}{},\x{0}{}\}$ 
adapted in the usual manner to $\Sigma$ and $\flow$, 
note $\lapse^{-1}=\nvec{0}{}$ is the inverse lapse 
and $\shift{i}{} = -\lapse \nvec{i}{}$ is the shift. 
Then, the linearly independent parts of a variation of the metric 
$\delta g$ are given by $\delta\h{}{ij}$ 
and $\delta\nvec{0}{}$, $\delta\nvec{i}{}$, 
or equivalently $\delta\lapse$, $\delta\shift{i}{}$. 
The stationary points of the Hamiltonian variational principle as defined by 
\EQ
0=\delta\H{}(\flow)/\delta\lapse ,\quad
0=\delta\H{}(\flow)/\delta\shift{i}{} ,\quad
0=\delta\H{}(\flow)/\delta\h{}{ij} 
\doneEQ
can be shown to yield, respectively, 
the constraint equations 
\EQ
\R{}{} +\K{j}{j}\K{i}{i} -\K{ij}{}\K{}{ij} =0 ,\quad
\D{j}\K{i}{j} - \D{i}\K{j}{j} =0
\doneEQ
and the evolution equation
\EQ
\lapse^{-1} \flowder{\K{ij}{}} = 
2\K{i}{l}\K{jl}{} -\K{ij}{}\K{l}{l} -\R{ij}{}
+\a{i}{}\a{j}{} +\D{i}\a{j}{}
+ 2\lapse^{-2} \shift{l}{}\K{l(i}{} \D{j)}\lapse 
\doneEQ
with 
\EQ
\lapse^{-1} \flowder{\h{ij}{}} = 
2\K{ij}{} 
+2\lapse^{-1} \D{(i}\shift{}{j)} 
\doneEQ
where $\D{i} =\h{i}{j}\covder{j}$ is the spatial derivative on $\Sigma$, 
$\flowder{}=\Lie{\flow}$ is the time derivative associated with $\flow$, 
$\R{ij}{}$ is the Ricci curvature tensor of $\h{ij}{}$ 
and $\R{}{}$ is its scalar curvature, 
while $\a{i}{}=\lapse^{-1}\D{i}\lapse$ is the spatial part of 
the acceleration $\covder{\n}\n$ of the hypersurface normal $\n$ on $\Sigma$. 

It is worth noting in these adapted coordinates that 
the geometrical vectors $\mcurv,\perpmcurv,\twist$
are expressed purely in terms of 
the Cauchy data $(\h{ij}{},\K{ij}{})$ on $\Sigma$
and the outward unit normal $\uvec{i}{}$ to $S$ in $\Sigma$ by 
\EQs
&&
\Hvec{i}{} = \trk(\u)\uvec{i}{} + \trk(\n) \lapse^{-1}\shift{i}{} ,\quad
\Hvec{0}{} = -\lapse^{-1}\trk(\n) ,
\\
&&
\perpHvec{i}{} = -\trk(\n)\uvec{i}{} - \trk(\u) \lapse^{-1}\shift{i}{} ,\quad
\perpHvec{0}{} = \lapse^{-1}\trk(\u) ,
\\
&&
\absmcurv=\absperpmcurv=\sqrt{|\trk(\u)^2-\trk(\n)^2|} ,
\doneEQs
and
\EQ
\wvec{i}{} = \uvec{l}{} \metric{}{ij} \K{jl}{} +\coSder{i}\chi . 
\doneEQ
Here 
$\trk(\n)= \metric{}{ij} \K{ij}{}$ 
and $\trk(\u)= \Sder{i}\uvec{i}{}$ 
are mean extrinsic curvatures of $S$,
while $\cosh\chi = \trk(\n)/\absmcurv$ 
and $\sinh\chi = \trk(\u)/\absmcurv$ 
determine the boost parameter $\chi$ relating $\n$ and $\perpmcurv$ 
at each point on $S$. 

The previous theorem is not intended to address the issue of whether 
the initial boundary value problem for the Einstein field equations 
is well-posed 
(\ie/ if there exist solutions $g$ satisfying the boundary conditions, 
initial conditions and constraints),
nor does it deal with the related issue of construction of 
a phase space for such solutions. 
Rather, the theorem simply provides a well-defined variational principle 
yielding the Einstein field equations in 3+1 form on spacelike hypersurfaces
$\Sigma$ with spatial boundary 2-surfaces $S=\partial\Sigma$. 
Its main application will be to give a covariant derivation of 
quasilocal gravitational quantities. 

On solutions of the Einstein field equations, 
the value of the Dirichlet Hamiltonian reduces to the surface integral 
\EQ
\H{\partial\Sigma}(\flow) = 
\frac{1}{8\pi}
\oint_{\partial\Sigma} \flow\cdot ( \perpmcurv +\twist_B +\refHdens{} ) \d{S} .
\doneEQ
The associated vector field $\P_B\equiv  \perpmcurv +\twist_B$ 
is called the gravitational {\it Dirichlet symplectic vector} \cite{paperI} 
and it is well defined just given the 2-surface $(\partial\Sigma,\sigma)$ 
along with the extrinsic geometry of $\partial\Sigma$ 
with respect to the boundary hypersurface $B$. 
Note, 
since the only vector available depending just 
on the boundary data $q$ and the 2-surface $\partial\Sigma$ is $\e{0}{}$, 
it follows that $\refHdens{}=\e{0}{} \refHdens{0}$
where $\refHdens{0}$ depends only on $\sigma$. 
Now consider the choice of flows $\flow=\e{0}{}$ and $\flow=\e{1}{}$ 
given by the adapted normal frame for $\partial\Sigma$ associated with $B$. 
The resulting surface integrals
\EQs
&& 
\H{\partial\Sigma}(\e{0}{}) 
= \frac{1}{8\pi} \oint_{\partial\Sigma} 
\e{0}{} \cdot (\perpmcurv +\refHdens{}) \d{S} 
= -\frac{1}{8\pi} \oint_{\partial\Sigma} ( \refHdens{0} + \trk(\e{1}{}) )\d{S} 
\\
&& 
\H{\partial\Sigma}(\e{1}{}) 
= \frac{1}{8\pi} \oint_{\partial\Sigma} \e{1}{} \cdot (\perpmcurv +\refHdens{}) \d{S} 
= -\frac{1}{8\pi} \oint_{\partial\Sigma} \trk(\e{0}{}) \d{S} 
\doneEQs
yield, respectively, Brown and York's expressions \cite{BrownYork2,canonical}
for quasilocal energy and quasilocal normal momentum 
(\ie/ outward-directed component of linear momentum). 
The term $\oint_{\partial\Sigma} \refHdens{0} \d{S}$ in the energy expression 
is called the reference energy. 
It is common to fix $\refHdens{0}$ 
by considering an isometric embedding of $(\partial\Sigma,\sigma)$ into 
a hyperplane $\Sigma\embed =\Rnum^3$ in Minkowski space $\Rnum^{3,1}$
and put $\refHdens{0} = \flow\embed \hook(\perpmcurv + \twist_B)\embed$
where $\flow\embed$ is identified with the vector $\e{0}{}\embed$ 
in the normal frame for $\partial\Sigma$
associated with the hyperplane $\Sigma\embed$. 
By Weyl's theorem \cite{weylembed}, 
such an embedding exists 
(and is unique up to isometry, \ie/ rigid motions of the 2-surface)
provided that $\partial\Sigma$ 
has positive intrinsic scalar curvature, $\scurvS>0$. 
This positivity implies $\partial\Sigma$ has the topology of a 2-sphere
since according to the Gauss-Bonnet theorem \cite{gaussbonnet},
$\oint \scurvS \d{S}= 4\pi (1-\genus)$
will be positive only if the topological genus 
(which measures the number of handles) of $\partial\Sigma$ is $\genus=0$. 
Because $\Sigma\embed$ is a hyperplane, note that
$\trk(\e{0}{}\embed)=0$
and hence $\perpmcurv\embed = \trk(\e{1}{}\embed) \e{0}{}\embed$.
This yields 
\EQ
-\refHdens{0}= \trk(\e{1}{}\embed) = \trk(\e{1}{})\eucl
\doneEQ
which is simply the extrinsic scalar curvature of $\partial\Sigma$
as a 2-surface embedded in Euclidean space $\Rnum^3$. 
Therefore the Brown-York quasilocal energy is given by 
\EQ\label{BYenergy}
\H{\partial\Sigma}(\e{0}{}) 
=\frac{1}{8\pi} \oint_{\partial\Sigma} 
( \trk(\e{1}{})\eucl -\trk(\e{1}{}) )\d{S} 
\equiv \qlE{BY}(\partial\Sigma,B)
\doneEQ
namely the difference in the 3-dimensional total mean curvature of 
$\partial\Sigma$
measured in a spacelike hypersurface orthogonal to $B$
versus a Euclidean reference space. 

Brown and York's expression for quasilocal angular momentum 
arises similarly from 
the choice of a flow $\flow=\phi$ 
given by a Killing vector on $\partial\Sigma$
(assuming the 2-surface admits a rotational symmetry, 
$\Lie{\phi} \metric{}{} =0$).
This yields 
\EQ\label{BYanglmom}
\H{\partial\Sigma}(\phi) 
= \frac{1}{8\pi}
\oint_{\partial\Sigma} \phi \cdot (\twist_B +\refHdens{}) \d{S} 
= \frac{1}{8\pi} \oint_{\partial\Sigma} 
{\e{1}{}\cdot\covder{\phi}\e{0}{}} \d{S} 
\equiv \qlJ{BY}_\phi(\partial\Sigma)
\doneEQ
with $\phi \cdot \twist_B = \e{1}{}\cdot\covder{\phi}\e{0}{}$
and $\phi \cdot \refHdens{} =0$
which follows from $\refHdens{}=\e{0}{} \refHdens{0}$. 

Thus $\perpmcurv$ is seen to have the role of 
a local rest-frame energy-momentum vector
that belongs to the normal space $\Tperpsurf$
and depends just on the extrinsic geometry of $\partial\Sigma$ in spacetime,
while $\twist_B$ has the role of a local angular-momentum vector
that belongs to the tangent space $\Tsurf$
and depends on the boundary hypersurface $B$ in spacetime. 
However, 
the quasilocal angular momentum \eqref{BYanglmom} is independent of $B$.
Specifically,
if the direction of $B$ orthogonal to $\partial\Sigma$ is boosted 
at points on $\partial\Sigma$,
then $\twist_B$ changes by a gradient $\covder{}\chi$
where the function $\chi$ on $\partial\Sigma$ represents the boost parameter.
Hence, since Killing vectors $\phi$ are divergence free, 
$\phi \cdot \twist_B$ changes just by 
$\phi \cdot \covder{}\chi =\Sder{}\cdot(\chi\phi)$
which is an irrelevant total divergence on $\partial\Sigma$
(\ie/ $\oint_S \phi \cdot \covder{}\chi \d{}S=0$ by Stokes' theorem).
In contrast the quasilocal energy \eqref{BYenergy}
has the quite unsatisfactory feature of 
depending on an arbitrary choice 
for the boundary hypersurface $B$ and Hamiltonian flow $\flow$. 
This dependence is typically phrased as a lack of invariance 
under boosts of the normal frame
$\{\e{0}{},\e{1}{}\}$, corresponding to a change of $B$. 
Moreover, both $B$ and $\flow$ are non-dynamical in the sense that 
the vector field $\flow$ on $M$ is independent of the spacetime metric $\g{}{}$
while the hypersurface $B$ is determined by Lie dragging of $S$ along
this vector field.
So, in such a Hamiltonian setting, 
$\flow$ and $B$ remain fixed under changes of the Cauchy data on $\Sigma$. 

But for applications in numerical relativity and in gluing of solutions,
dynamical boundaries that depend on the time evolution of the 
gravitational field variables on $\Sigma$
are of interest. 
Thus the aim now will be to consider dynamical time-flow vector fields $\flow$
(called ``live'' fields in numerical relativity)
and associated timelike boundaries $B$ that are allowed to change with $g$.

\subsection{ Mean curvature Hamiltonians and flows }

A natural proposal in terms of the geometry of spacelike 2-surfaces $S$ 
in spacetime 
is to consider a dynamical time-flow defined by 
the dual mean curvature vector
\EQ
\flow|_S = \absmcurv^{-1} \perpmcurv \equiv \unitperpmcurv . 
\doneEQ
For 2-surfaces $S=\partial\Sigma$ that are regular, 
this flow is timelike or null. 
The Hamiltonian analysis carried out in \Ref{paperII}
gives the following result. 

Theorem~3.2:
Let boundary conditions on $g$ at a convex-boundary $\partial\Sigma$ 
be defined by fixing 
the mean curvature timelike-flow $\unitperpmcurv$ 
and 2-surface metric $\sigma$. 
Then a Hamiltonian for the 3+1 form of the Einstein field equations on $\Sigma$
is given by 
\EQ\label{boundaryH}
\H{}(\unitperpmcurv) = 
\H{\rm ADM}(\xi)
+\frac{1}{8\pi} \oint_{\partial\Sigma} (\refHdens{\perp} -\absmcurv) \d{S}
\doneEQ
where $\refHdens{\perp}$ depends only on $\sigma$,
and where $\xi$ denotes a smooth extension of $\flow|_S=\unitperpmcurv$
from $\partial\Sigma$ into $\Sigma$. 

A natural choice of $\refHdens{\perp}$ is $\abseuclmcurv$ 
with $(S,\sigma)$ embedded into a hyperplane in Minkowski space. 
When evaluated on solutions of the field equations, 
the Hamiltonian \eqref{boundaryH}
yields a geometrical quantity, 
the difference of the 4-dimensional total mean curvature of $S$ 
in spacetime and in Minkowski space, 
\EQ
\H{\partial\Sigma}(\unitperpmcurv) = 
\frac{1}{8\pi} \oint_{\partial\Sigma} (\abseuclmcurv -\absmcurv) \d{S}
\equiv \qlE{MC}(S)
\label{mcurvqlm}
\doneEQ
which depends on just $S$, $\sigma$, and $g$. 
Clearly, 
its interpretation relative to the Brown-York quantities 
is that of a ``rest mass'', 
analogous to the relation between ADM energy-momentum and mass. 
Accordingly the quasilocal quantity \eqref{mcurvqlm} 
will be referred to as the {\it mean curvature mass}. 
An equivalent definition but in a less covariantly geometrical form 
appears in work of 
Kijowski \cite{Kijowski}, Lau \cite{Lau}, Epp \cite{Epp}, 
Liu and Yau \cite{LiuYau}, and others \cite{unpublished}. 

There is an interesting generalization of the previous theorem, 
leading to a family of geometric quasilocal mass definitions that include 
as special cases both the mean curvature mass and the Hawking mass. 
Take the Hamiltonian flow at $S$ to be 
\EQ
\flow|_S = c(\sigma) \absmcurv^{n-1} \perpmcurv 
\doneEQ
in terms of a parameter $n$, 
where $c(\sigma)$ depends only on $\sigma$. 
Then the following result will be proved later. 

Theorem~3.3:
Let the 2-surface metric $\sigma$ 
and the mean curvature timelike-flow vector $\absmcurv^n \unitperpmcurv$ 
be fixed boundary data at a convex-boundary $\partial\Sigma$. 
Under these boundary conditions on $g$, 
there exists a Hamiltonian for the 3+1 form of 
the Einstein field equations on $\Sigma$, 
given by the ADM Hamiltonian $\H{\rm ADM}(\xi)$ 
plus the boundary term 
$\H{\partial\Sigma}^{(n)}(\unitperpmcurv) = 
\frac{1}{8\pi}
\oint_{\partial\Sigma} 
(\refHdens{\perp}(\sigma) -\frac{1}{n+1} c(\sigma) \absmcurv^{n+1}) \d{S}$
where $\refHdens{\perp}$ depends only on $\sigma$,
and where the time-flow $\xi$ 
is a smooth extension of $\flow|_S$ into $\Sigma$. 

So that $\H{\partial\Sigma}^{(n)}(\unitperpmcurv)$ 
has physical units of energy, 
put $c(\sigma)=c_0 \sqrt{A}^n$ with $c_0=\const$
and $A= \oint_S \d{S}$ denoting the area of $S$, 
and also factor out $c(\sigma)/(n+1)$ from $\refHdens{\perp}(\sigma)$. 
These choices yield
\EQ\label{nthmcurvbt}
\H{\partial\Sigma}^{(n)} (\unitperpmcurv) = 
\frac{c_0}{8\pi(n+1)} \sqrt{A}^n 
\oint_{\partial\Sigma} (\refHdens{\perp}(\sigma) - \absmcurv^{n+1}) \d{S}
\doneEQ
where $\refHdens{\perp}(\sigma)$ now is required to have 
the same physical units as $\absmcurv^{n+1}$.
For $n=0$ the expression $\H{\partial\Sigma}^{(0)}(\unitperpmcurv)$ 
reduces to the mean curvature mass \eqref{mcurvqlm} 
if $\refHdens{\perp}(\sigma)= \abseuclmcurv$
is given by the isometric embedding of $(S,\sigma)$ as before, 
which assumes $S$ is a topological 2-sphere 
with positive scalar curvature.
On the other hand for $n=1$, 
$\H{\partial\Sigma}^{(1)}(\unitperpmcurv)$ 
is proportional to the Hawking mass 
\EQ\label{hawkingqlm}
\qlE{H}(S) \equiv
\frac{\sqrt{A}}{16\pi} \oint_{\partial\Sigma} (4\scurvS -\absmcurv^2) \d{S}
\doneEQ
if $\refHdens{\perp}(\sigma)/4= \scurvS$ is chosen equal to 
the scalar curvature of $S$. 
Because no embedding is involved in this case, 
there are no restrictions needed on the topology or scalar curvature for $S$
and thus an equivalent choice for $\refHdens{\perp}(\sigma)$ is 
$16\pi (1-\genus)/A$ 
since, by the Gauss-Bonnet theorem, 
$\oint \scurvS \d{S}= 4\pi (1-\genus)$
where $\genus$ is the topological genus of $S$
(\ie/ $\genus=0$ is 2-sphere, $\genus=1$ is a torus, \etc/).

A further natural generalization is obtained 
by taking the Hamiltonian flow to be along the direction of 
the Dirichlet symplectic vector $\P = \perpmcurv +\twist$
associated to the mean curvature frame $\{\perpmcurv,\mcurv\}$ of
$\partial\Sigma$. 
So consider
\EQ\label{Pflow}
\flow|_S = c(\sigma) \absP^{n-1} \P
\doneEQ
where $\absP = \sqrt{|\sqP|}$, 
and $c(\sigma)$ depends only on $\sigma$. 
Observe that 
\EQ
\sqP = \perpmcurv^2 +\twist^2 ,\quad
\mcurv^2 = -\perpmcurv^2 
\doneEQ
and so 
\EQ
\sqP \leq 0 \eqtext{ iff } \absmcurv \geq |\twist| \geq 0 . 
\doneEQ
From the expressions for the Brown-York quasilocal quantities, 
on a convex-regular 2-surface $S$
the scalar mean curvature $\absmcurv$ may be interpreted 
as a ``mass density''
and the mean curvature twist $\twist$ 
as an ``angular momentum density'' vector. 
This motivates introducing the local ``energy'' condition
$\absmcurv \geq |\twist|$ on a spacelike 2-surface $S$
so that $\P$ is a timelike or null vector 
at every point on $S$. 
Notice if this condition holds on $S$
then it implies $S$ is a convex-regular 2-surface. 

Definition~3.4:
A 2-surface $S$ will be called {\it twist-free} 
if $\twist(\mcurv)=0$ holds everywhere on $S$. 

Note if a convex-regular 2-surface is twist-free 
then its mean curvature twist vanishes, 
$\twist=\absmcurv^{-2}\twist(\mcurv)=0$.
Consequently the full expansion of such a 2-surface vanishes 
in the direction of the dual mean curvature vector
\EQ
\Lie{\perpmcurv} \vol{}{}(S) = 0 . 
\doneEQ
The following result will be proved later. 

Theorem~3.5:
Suppose a 2-surface $S=\partial\Sigma$ 
satisfies the local energy condition 
$\sqP < 0$ (implying $S$ is a convex-boundary). 
Let the 2-surface metric $\sigma$ 
and the Dirichlet vector $\absP^{n-1} \P$ 
be fixed boundary data at $\partial\Sigma$. 
Take $\flow$ to be a smooth extension of the symplectic flow 
\eqref{Pflow} into $\Sigma$. 
Under these boundary conditions on $g$, 
a Hamiltonian for the 3+1 form of the Einstein field equations on $\Sigma$ 
is given by the ADM Hamiltonian $\H{\rm ADM}(\xi)$
plus the boundary term 
$\H{\partial\Sigma}^{(n)}(\unitP) = 
\frac{1}{8\pi(n+1)} 
\oint_{\partial\Sigma} 
c(\sigma) (\refHdens{\perp}(\sigma) -\absP^{n+1}) \d{S}$
where $\refHdens{\perp}$ depends only on $\sigma$
(with the factor $c(\sigma)/(n+1)$ having been scaled out for convenience). 

As earlier, 
it is natural to put $c(\sigma)=c_0 \sqrt{A}^n$ 
so $\H{\partial\Sigma}^{(n)}(\unitP)$ will have physical units of energy,
while $\refHdens{\perp}(\sigma)$ shares the same units as $\absmcurv^{n+1}$,
so thus 
\EQ
\H{\partial\Sigma}^{(n)}(\unitP) = 
\frac{c_0}{8\pi(n+1)} \sqrt{A}^n
\oint_{\partial\Sigma} (\refHdens{\perp}(\sigma) -\absP^{n+1}) \d{S} . 
\doneEQ

Some remarks on the geometrical, spacetime meaning of the boundary conditions
in theorems~3.3 and~3.5 are appropriate. 
Observe that the boundary data $\sigma$ 
and $\absP^{n-1} \P$ or $\absmcurv^{n-1} \perpmcurv{}{}$ 
together define a family of flows of the 2-surface $S$ in spacetime, 
each of which locally generates some timelike hypersurface $B$ 
when $S$ satisfies the local energy condition $\sqP < 0$. 
The component of these flows orthogonal to $S$ is given by 
the direction of the timelike mean curvature vector $\perpmcurv$. 
This yields, geometrically, a mean curvature flow of 2-surfaces
produced by $S$ being Lie dragged along the direction $\unitperpmcurv$, 
with the flows differing just in the speed, which depends on $n$. 
In the case $n=0$, 
it follows that the boundary conditions in the Hamiltonian 
have the geometrical content of fixing the intrinsic metric on 
the resulting timelike hypersurface (\ie/ the image of $S$ in the flow)
in spacetime. 
(An obvious analytical question of interest is
how long the flow will remain smooth and timelike, 
and what will happen to the intrinsic geometry of the 2-surface in the flow.)

The tangential component of the above family of flows on $S$
is proportional to the mean curvature twist vector $\twist$. 
This flow is less interesting geometrically than 
is the timelike mean curvature flow, nevertheless
it is still relevant in a Hamiltonian setting as follows.

Theorem~3.6:
Let the boundary data at a convex-boundary $S=\partial\Sigma$ 
consist of 
the 2-surface metric $\sigma$ 
and mean curvature twist $\twist$. 
Consider a Hamiltonian flow $\flow$ given by a smooth extension of 
$\flow|_S= c(\sigma)\twist$ on $S$ into $\Sigma$. 
Then the ADM Hamiltonian $\H{\rm ADM}(\xi)$
yields a well-defined variational principle
under the addition of the boundary term 
$\H{\partial\Sigma}(\twist) = 
\frac{1}{16\pi} \oint_{\partial\Sigma} 
c(\sigma)( \refHdens{\parallel}(\sigma) +|\twist|^2 ) \d{S}$. 

If $\twist$ is interpreted as a rotational covector field on $S$ 
then $\H{\partial\Sigma}(\twist)$ will have physical units of 
squared angular momentum 
by putting $c(\sigma)=c_0 A^{2}$. 

A slight generalization of the previous theorem arises from 
using the Hamiltonian flow $\flow|_S= c(\sigma)\absmcurv^n\twist$,
together with the boundary condition that $\absmcurv$ is fixed
in addition to $\twist$ at $S=\partial\Sigma$. 
The resulting boundary term in the Hamiltonian is given by 
\EQ
\H{\partial\Sigma}^{(n)}(\twist) = 
\frac{c_0}{16\pi}\sqrt{A}^{4+n} \oint_{\partial\Sigma} 
(\refHdens{\parallel}(\sigma) +\absmcurv^n |\twist|^2) \d{S} 
\doneEQ
where $\H{\partial\Sigma}^{(n)}(\twist)$ is defined to have physical units of 
squared angular momentum. 

Proofs of theorems~3.2, 3.3, 3,5, 3.6 will be given next.

\subsection{ Noether charge analysis for 
Hamiltonians with mean curvature time flow }

The Noether charge formalism is summarized as follows \cite{paperI}. 
Consider a general diffeomorphism-invariant field theory
described by a Lagrangian 4-form $\L(\phi)$, 
for dynamical fields $\phi$ on spacetime $M$. 
The symplectic potential is a 3-form $\Theta(\phi,\delta\phi)$ 
related to the equations of motion, $\delta\L/\delta\phi= \E{}(\phi)=0$, 
through the variation of the Lagrangian,
\EQ
\delta \L(\phi)=
\E{}(\phi) \delta \phi  + d \Theta (\phi,\delta\phi) 
\doneEQ
for arbitrary variations $\delta\phi$. 
The symplectic current 3-form $\omega$ is defined by
\EQ
\omega(\delta_1\phi,\delta_2\phi)=
\delta_1\Theta(\phi,\delta_2\phi)- \delta_2\Theta(\phi,\delta_1\phi) 
\doneEQ
and remains unchanged under addition of any exact 4-form to $\L$. 
Integration of $\omega$ over a spacelike hypersurface $\Sigma$
gives the presymplectic form $\Omega = \int_\Sigma \omega$,
which encodes the symplectic structure of the field theory
in a covariant manner \cite{Witten,WaldLee}. 

Suppose a flow is given on $M$ by a vector field $\xi$,
which will be allowed to depend on $\phi$. 
If the presymplectic form is a total variation on shell
(\ie/ when $\E{}(\phi)=0$)
\EQ
\Omega(\phi, \delta\phi, \pounds_\xi \phi)
= \int_\Sigma\omega(\phi, \delta\phi,\pounds_\xi\phi)
=\delta  H(\xi)
\doneEQ
for some functional $H(\xi)=\int_\Sigma \Hdens(\phi,\xi) \d{\Sigma}$, 
then $H(\xi)$ is conserved along $\xi$, 
\ie/ $\Lie{\xi} H(\xi)=0$. 
This functional $H(\xi)$ is called the Hamiltonian conjugate to $\xi$.

Associated with the flow $\xi$ is a Noether current 3-form $J$ defined by
\EQ
J(\phi,\xi)=\Theta(\phi,{\rm\pounds}_\xi\phi)-\xi\hook \L(\phi)
\doneEQ
where $\Lie{\xi}$ denotes the Lie derivative 
and $\xi\hook$ is the interior product. 
Because $\delta_\xi \phi \equiv \Lie{\xi}\phi$ 
generates a diffeomorphism symmetry of the Lagrangian,
\EQ
\delta_\xi\L(\phi) = \d{}(\xi\hook\L(\phi) = \Lie{\xi}\L(\phi) , 
\doneEQ
the Noether current is conserved, \ie/ closed on shell, 
since $\d{}J(\phi,\xi)=-\E{}(\phi)\Lie{\xi}\phi=0$
when $\E{}(\phi)=0$. 
Consequently, as the conservation holds for any flow $\xi$,
it can be shown by an application of a homotopy integral formula
\cite{Anderson,Wald}
with respect to $\xi$
that, on shell, the Noether current is exact.
Hence there is a 2-form $Q$ (the Noether potential) given by 
\EQ
J(\phi,\xi)=\d{} Q(\phi,\xi) . 
\doneEQ
Under variations of $\phi$ with $\xi$ kept fixed, $\delta\xi=0$, 
the Noether current satisfies 
\EQ
\delta J(\phi,\xi)= 
\omega(\phi,\delta\phi, {\rm\pounds}_\xi\phi) 
+ \d{} (\xi\hook \Theta (\phi,\delta\phi ) ) . 
\doneEQ
This implies, on shell, 
\EQ
\Omega(\phi, \delta\phi, \pounds_\xi \phi)
= \delta\int_\Sigma J(\phi,\xi) 
- \int_{\partial\Sigma} \xi\hook \Theta (\phi,\delta\phi ) . 
\doneEQ
So, in order to have a well-defined Hamiltonian, 
the boundary term needs to be a total variation, 
namely there must exist a 2-form $B(\phi,\xi)$ such that 
at the boundary $\partial\Sigma$ 
\EQ
\xi\hook \Theta(\phi,\delta\phi)=\delta B(\phi,\xi) 
\doneEQ
holding on shell.
Then the on-shell Hamiltonian will be given by 
\EQ
H(\xi)=\oint_{\partial\Sigma} Q(\phi,\xi) - B(\phi,\xi) .
\doneEQ

Now take the Hilbert Lagrangian for General Relativity, 
using an orthonormal frame 1-form field $\e{}{a}$ 
for the gravitational field variable 
\cite{Nester2},
\EQ
\L 
= \textstyle{1\over 4!} \scurv \vol{abcd}{}
\e{}{a}\wedge\e{}{b}\wedge\e{}{c}\wedge\e{}{d}
= \curv{}{ab}\wedge \vole{ab}{}
\doneEQ
where 
$\vole{ab}{}= 
\frac{1}{2} \vol{abcd}{}\e{}{c}\wedge\e{}{d}$, 
$\curv{}{ab}=\d{}\conx{}{ab}+\conx{c}{a} \wedge \conx{}{bc}$ 
is the curvature 2-form,
$\conx{b}{a}$ is the frame connection 1-form determined by 
$\d{}\e{}{a} = \conx{b}{a}(e) \wedge \e{}{b}$, 
$\scurv = \frac{1}{2}(\e{a}{}\wedge \e{b}{})\hook \curv{}{ab}$
is the scalar curvature (trace of the Ricci tensor), 
using the coframe $\e{a}{}$ of $\e{}{a}$, \ie/
$\e{a}{}\hook \e{}{b} =\id{a}{b}$. 
(Note $\e{a}{}\wedge \e{b}{}$ denotes the bi-vector dual to 
the 2-form $\e{}{a}\wedge\e{}{b}$.)
For simplicity 
the standard factor of $1/16\pi$ has been dropped in the Lagrangian. 

After use of the fact that $\conx{b}{a}(e)$ satisfies 
$\delta\L/\delta\conx{b}{a} =0$
(implying the connection is metric compatible), 
an arbitrary variation of the Lagrangian gives
\EQ
\delta \L
= \vol{abcd}{} \curv{}{ab}\wedge \e{}{c} \wedge \delta \e{}{d}
+ d( \delta \conx{}{ab}(e) \wedge \vole{ab}{} )
\doneEQ
which yields
\EQ
\Theta(\e{}{},\delta\e{}{})
=\delta \conx{}{ab}(e) \wedge \vole{ab}{} 
\doneEQ
and 
\EQ
0= \E{a}(e) = \delta\L/\delta\e{}{a} 
= \e{}{b}\wedge \curv{}{cd} \vol{bcda}{}
\doneEQ
where 
$\frac{1}{2} *\E{a}(e) = \e{a}{}\hook(\ricci-\frac{1}{2} g\scurv)$
are the Einstein field equations on $\e{}{a}$. 

The Noether current 3-form is given by \cite{paperI,paperII}
\EQ
J(\e{}{},\xi) = \d{} Q(\e{}{},\xi) +(\flow\hook\e{}{a})\E{a}(e)
\doneEQ
where 
\EQ
Q(\e{}{},\xi)= (\xi\hook \conx{}{ab}(e)) \vole{ab}{}
\doneEQ
is the Noether charge 2-form potential. 

To prove theorem~3.5, 
consider a hypersurface $\Sigma$ with a spacelike boundary 2-surface $S$ 
on which the local energy condition $\P\cdot\P<0$ holds
and with the following boundary conditions:
Let a smooth extension of 
\EQ
\xi|_S = \absP^n \P
\doneEQ
into $\Sigma$ define the time-flow $\xi$,
with $\absP = \sqrt{-\P \cdot \P}$.
One boundary condition will be to fix this time-flow vector 
at the 2-surface
\EQ
\delta \xi |_S =0 . 
\label{bc1}
\doneEQ
Another boundary condition will be to fix the intrinsic metric of 
the 2-surface 
\EQ
\delta \e{\indS{a}}{} |_S=0 ,
\qquad \indS{a}=2,3 
\label{bc2}
\doneEQ
where $\e{\indS{a}}{}$ denotes the tangential frame of $S$. 
In addition, it will be convenient to adapt the normal frame 
to the extrinsic geometry of $S$ by putting 
\EQ
\e{0}{} |_S = \absmcurv^{-1} \perpmcurv ,\quad
\e{1}{} |_S = \absmcurv^{-1} \mcurv 
\label{bc3}
\doneEQ
\ie/ the mean curvature frame. 
Hereafter we write the coframe for $\coTM|_S$
satisfying equations \eqrefs{bc2}{bc3}
as $\frame{}{a}$. 
Then using boundary conditions \eqrefs{bc1}{bc2}, we have 
\EQs
\oint_S \xi\hook\Theta(\frame{}{},\delta\frame{}{}) &&
= \oint_S \xi\hook( 
\delta \conx{}{ab}(\frame{}{}) \wedge \volframe{ab}{} )
\nonumber\\&&
= \oint_S \delta( 
\xi\hook( \conx{}{ab}(\frame{}{}) \wedge \volframe{ab}{} ))
+ (\xi\hook \delta \frame{}{c}) 
\vol{abcd}{} \conx{}{ab}(\frame{}{}) \wedge\frame{}{d} . 
\label{Theta}
\doneEQs
Note that the following identity holds on the
2-surface $S$ for any vector field $V^a \e{a}{}$:
\EQ
\vol{abcd}{} V^c \conx{}{ab}(e) \wedge \e{}{d}
= (- V^0 \trk(\e{1}{}) - V^1 \trk(\e{0}{}) 
+ V^\indS{a} \twist(e)_\indS{a} ) 
\vol{\indS{c}\indS{d}}{}\ \e{}{\indS{c}}\wedge \e{}{\indS{d}}
\doneEQ
which reduces to 
\EQ
\label{mainid}
\vol{abcd}{} V^c \conx{}{ab}(\frame{}{}) \wedge \frame{}{d}
= ( -V^0 \absmcurv + V^\indS{a} \twist_\indS{a} ) 
\vol{\indS{c}\indS{d}}{} \frame{}{\indS{c}}\wedge \frame{}{\indS{d}}
\doneEQ
in the adapted coframe $\frame{}{a}$. 

In the last term in the integral \eqref{Theta}, 
this identity yields 
\EQs
\oint_S (\xi\hook \delta \frame{}{c}) \vol{abcd}{} 
\conx{}{ab}(\frame{}{}) \wedge \frame{}{d} 
&&
= \oint_S 
\xi\hook( -\delta\frame{}{0} \absmcurv
+ \delta\frame{}{\indS{a}} \twist_\indS{a} )
\vol{\indS{c}\indS{d}}{} \frame{}{\indS{c}}\wedge \frame{}{\indS{d}}
\nonumber\\&&
= \oint_S 
\delta ( 
-\xi\hook\frame{}{0} \absmcurv
+ \xi\hook\frame{}{\indS{a}} \twist_\indS{a} ) 
\vol{\indS{c}\indS{d}}{} \frame{}{\indS{c}}\wedge \frame{}{\indS{d}}
\nonumber\\&&\qquad 
+ ( \xi\hook\frame{}{0} \delta\absmcurv
- \xi\hook\frame{}{\indS{a}} \delta\twist_\indS{a} ) 
\vol{\indS{c}\indS{d}}{} \frame{}{\indS{c}}\wedge \frame{}{\indS{d}}
\nonumber\\&& 
= \oint_S ( -\delta \absP^{n+2} 
+ \absP^n (\absmcurv\delta\absmcurv -\twist^\indS{a} \delta\twist_\indS{a}) )
\vol{\indS{c}\indS{d}}{} \frame{}{\indS{c}}\wedge \frame{}{\indS{d}}
\nonumber\\&& 
= -\smallfrac{n+1}{n+2} \oint_S \delta \absP^{n+2} 
\vol{\indS{c}\indS{d}}{} \frame{}{\indS{c}}\wedge \frame{}{\indS{d}}
\nonumber\\&&
=\smallfrac{n+1}{n+2} \oint_S
\delta( (\xi\hook \frame{}{c}) 
\vol{abcd}{} \conx{}{ab}(\frame{}{}) \wedge \frame{}{d} ) , 
\doneEQs
and hence 
\EQs
\oint_S \xi\hook\Theta(\frame{}{},\delta\frame{}{}) && 
= \oint_S \delta (\xi\hook (\conx{}{ab}(\frame{}{}) \wedge \volframe{ab}{} ))
+ \smallfrac{n+1}{n+2} \delta( (\xi\hook \frame{}{c}) 
\vol{abcd}{} \conx{}{ab}(\frame{}{}) \wedge \frame{}{d} )
\nonumber\\&& 
=\oint_S \delta B(\frame{}{},\xi)
\doneEQs
where
\EQ
B(\frame{}{},\xi) = 
\xi\hook (\conx{}{ab}(\frame{}{}) \wedge \volframe{ab}{})
+ \smallfrac{n+1}{n+2} (\xi\hook \frame{}{c}) 
\vol{abcd}{} \conx{}{ab}(\frame{}{}) \wedge \frame{}{d} . 
\doneEQ
Thus we obtain 
\EQ
\Omega(e,\delta e,\Lie{\xi}e)
= \int_\Sigma\omega(e,\delta e,\Lie{\xi}e) 
= \int_\Sigma \delta( J(\frame{}{},\xi) -d B(\frame{}{},\xi) )
=\delta \H{}(\xi)
\doneEQ
where 
\EQ
\label{bt}
\H{}(\xi) =
\int_\Sigma (\xi\hook\frame{}{a}) \E{a}(e)
+ \oint_{S} Q(\frame{}{},\xi)- B(\frame{}{},\xi) 
\doneEQ
is the off-shell Hamiltonian. 
Its boundary term is given by 
\EQs
\H{\partial\Sigma}(\xi) 
&=&
\oint_S Q(\frame{}{},\xi) - B(\frame{}{},\xi)
\nonumber\\
&=&
\oint_S 
( \xi\hook \conx{}{ab}(\frame{}{}) ) \volframe{ab}{} 
- \xi\hook (\conx{}{ab}(\frame{}{}) \wedge \volframe{ab}{})
-\smallfrac{n+1}{n+2} (\xi\hook \frame{}{c}) 
\vol{abcd}{} \conx{}{ab}(\frame{}{}) \wedge \frame{}{d} 
\nonumber\\
&=&
\oint_S 
(1-\smallfrac{n+1}{n+2}) 
(\xi\hook \frame{}{c}) 
\vol{abcd}{} \conx{}{ab}(\frame{}{}) \wedge \frame{}{d} 
\nonumber\\
&=&
\smallfrac{1}{n+2} \oint_S \absP^{n} (-\absmcurv^2 +|\twist|^2)
\vol{\indS{c}\indS{d}}{} \frame{}{\indS{c}}\wedge \frame{}{\indS{d}}
\nonumber\\
&=&
-\smallfrac{2}{n+2} \oint_S \absP^{n+2} \d{S}
= \smallfrac{2}{n+2} \oint_S \xi\cdot\P \d{S}
\doneEQs
with 
$\d{S} =
{1\over 2}\vol{\indS{c}\indS{d}}{} 
\frame{}{\indS{c}} \wedge \frame{}{\indS{d}} 
= \vol{}{}(S)$
denoting the 2-surface area element. 
The volume term in $\H{}(\xi)$ is simply given by $16\pi$ times 
the ADM Hamiltonian \eqref{ADMH},
which vanishes on-shell. 

This establishes theorem~3.5.
Theorems~3.3 and~3.6 can be proved by a similar analysis.

\subsection{ Analysis of null flows and Hamiltonians }

It is interesting to generalize the previous results to consider
mean curvature null flows given by 
\EQ
\flow|_S= c(\sigma)\absmcurv^n \mcurv_+
\doneEQ
where $\mcurv_+ = \frac{1}{2}(\perpmcurv+\mcurv)$ is
the outward mean curvature null vector 
at a boundary 2-surface $S=\partial\Sigma$
which will not be assumed to be convex hereafter. 

Theorem~3.7:
Let the 2-surface metric $\sigma$ 
and scaled mean curvature null flow vector $\absmcurv^n \mcurv_+$
be fixed boundary data at $\partial\Sigma$. 
Then a Hamiltonian for the 3+1 form of the Einstein field equations
is given by the ADM Hamiltonian $\H{\rm ADM}(\xi)$ plus the boundary term 
\EQ\label{nthnullmcurvbt}
\H{\partial\Sigma}^{(n)}(\mcurv_+) = 
\frac{c(\sigma)}{8\pi(n+2)} \oint_{\partial\Sigma} 
( \refHdens{+}(\sigma) -\absmcurv^{n+2}) \d{S}
\doneEQ
where $\refHdens{+}$ depends only on $\sigma$,
and where $\xi$ is a smooth extension of 
$\flow|_S$ into $\Sigma$. 

The boundary term here is almost the same as the one in theorem~3.3
for the timelike flow $\flow|_S= c(\sigma)\absmcurv^{n-1} \perpmcurv$,
apart from a trivial change just in the overall numerical factor. 
In contrast the boundary data in theorem~3.7 has a different
geometrical meaning compared to theorem~3.3 as it yields 
a family of geometric null flows of the 2-surface $S$ in spacetime,
given by Lie dragging $S$ along the null mean curvature direction $\mcurv_+$,
locally generating a null hypersurface $B$. 

This null flow is nontrivial for $n>0$ provided $S$ is a convex-boundary
(in which case $\absmcurv>0$). 
For $n=0$, $S$ need not be convex to define the flow, 
since $\flow|_S= \mcurv_+$
is a well-defined null vector for any spacelike 2-surface $S$. 
Consequently, the $n=0$ case of theorem~3.7 is applicable to 
marginally outward trapped 2-surfaces and apparent horizons,
given by $\mcurv_+=\mcurv=\perpmcurv$ on $S$, and $\absmcurv=0$ on $S$,
respectively. 

The proof of theorem~3.7 will now be outlined.
As in the previous theorems, under the boundary conditions \eqrefs{bc1}{bc2}
we have 
\EQ
\oint_S \xi\hook\Theta(\frame{}{},\delta\frame{}{}) 
= \oint_S \delta( 
\xi\hook( \conx{}{ab}(\frame{}{}) \wedge \volframe{ab}{} ))
+ (\xi\hook \delta \frame{}{c}) 
\vol{abcd}{} \conx{}{ab}(\frame{}{}) \wedge\frame{}{d}  
\label{nullTheta}
\doneEQ
where $\frame{a}{}$ now denotes a null frame adapted to the 
null mean curvature vectors $\frame{\pm}{}|_S = \mcurv_\pm$.
The main identity \eqref{mainid} thus becomes
\EQ
\label{nullmainid}
\vol{abcd}{} V^c \conx{}{ab}(\frame{}{}) \wedge \frame{}{d}
= ( V^+ + \frac{1}{2}\absmcurv^2 V^- \mp V^\indS{a} \twist^\pm_\indS{a} ) 
\vol{\indS{c}\indS{d}}{} \frame{}{\indS{c}}\wedge \frame{}{\indS{d}}
\doneEQ
for any vector field $V^a\e{a}{}$ at $S$. 
Hence the last term in the integral \eqref{nullTheta} reduces to
\EQs
\oint_S (\xi\hook \delta \frame{}{c}) \vol{abcd}{} 
\conx{}{ab}(\frame{}{}) \wedge \frame{}{d} 
&&
= \oint_S 
\xi\hook( \delta\frame{}{+} +\smallfrac{1}{2}\absmcurv^2 \delta\frame{}{-} )
\vol{\indS{c}\indS{d}}{} \frame{}{\indS{c}}\wedge \frame{}{\indS{d}}
\nonumber\\&&
= \oint_S 
\smallfrac{1}{2}\delta ( \absmcurv^2 \xi\hook\frame{}{-} )
\vol{\indS{c}\indS{d}}{} \frame{}{\indS{c}}\wedge \frame{}{\indS{d}}
-\smallfrac{1}{2}\xi\hook\frame{}{-} \delta(\absmcurv^2)
\vol{\indS{c}\indS{d}}{} \frame{}{\indS{c}}\wedge \frame{}{\indS{d}}
\nonumber\\&& 
= \oint_S ( -\delta \absmcurv^{n+2} + \absmcurv^n \delta\absmcurv^2 )
\smallfrac{1}{2}
\vol{\indS{c}\indS{d}}{} \frame{}{\indS{c}}\wedge \frame{}{\indS{d}}
\nonumber\\&& 
= -\smallfrac{n}{n+2} \oint_S \delta \absmcurv^{n+2} \smallfrac{1}{2}
\vol{\indS{c}\indS{d}}{} \frame{}{\indS{c}}\wedge \frame{}{\indS{d}}
\nonumber\\&&
=\smallfrac{n}{n+2} \oint_S 
\delta( \smallfrac{1}{2} (\xi\hook \frame{}{c}) 
\vol{abcd}{} \conx{}{ab}(\frame{}{}) \wedge \frame{}{d} ) .
\doneEQs
As a result, the boundary term in the Hamiltonian \eqref{bt} is given by 
\EQs
\H{\partial\Sigma}(\xi) 
&=&
\oint_S Q(\frame{}{},\xi) - B(\frame{}{},\xi)
\nonumber\\
&=&
\oint_S 
\smallfrac{1}{2}(1-\smallfrac{n}{n+2}) 
(\xi\hook \frame{}{c}) 
\vol{abcd}{} \conx{}{ab}(\frame{}{}) \wedge \frame{}{d} 
\nonumber\\
&=&
-\smallfrac{2}{n+2} \oint_S \absmcurv^{n+2} \d{S}
= \smallfrac{4}{n+2} \oint_S \xi\cdot\P \d{S}
\doneEQs
which completes the proof. 
One remark concerning this boundary Hamiltonian is that 
since $\flow$ is a null vector the expression $\flow\cdot\P$
depends only on the normal part of the symplectic vector $\P$,
namely $\perpmcurv$, and thus is well-defined even when $S$ is not convex. 

Finally, it is worth stating that theorem~3.6 can be easily generalized
(with a similar proof) by considering 
the null mean curvature twist $\twist^\pm$
as a rotational Hamiltonian flow,
which is applicable to marginally outward trapped surfaces 
and apparent horizons.

Theorem~3.8:
For a Hamiltonian flow $\flow$ given by a smooth extension of 
$\flow|_S= c(\sigma)\absmcurv^n\twist^\pm$ on $S$ into $\Sigma$,
let the boundary data at $S=\partial\Sigma$ consist of 
the 2-surface metric $\sigma$ 
and null mean curvature twist $\twist^\pm$
in addition to the scalar mean curvature $\absmcurv$ for $n\neq 0$. 
Then the ADM Hamiltonian $\H{\rm ADM}(\xi)$
yields a well-defined variational principle
under the addition of the boundary term 
$\H{\partial\Sigma}^{(n)}(\twist^\pm) = 
\frac{1}{16\pi} \oint_{\partial\Sigma} 
c(\sigma)( \refHdens{\pm}(\sigma) +\absmcurv^n |\twist^\pm|^2 ) \d{S}$
where $\refHdens{\pm}$ depends only on $\sigma$. 

Note $\H{\partial\Sigma}^{(n)}(\twist^\pm)$ will have physical units of 
squared angular momentum if 
$c(\sigma)=c_0 \sqrt{A}^{4+n}$ as before.

\section{ Quasilocal mean-curvature quantities }
\label{quasilocal}

The various boundary Hamiltonians $\H{\partial\Sigma}$ coming
from theorems~3.3 and 3.5~to~3.8 
can be naturally studied as geometric quasilocal quantities
defined for spacelike 2-surfaces in any spacetime, 
and not just solutions of the Einstein field equations. 
This will yield quasilocal mass-energy definitions 
as well as quasilocal angular momentum definitions
underpinned by Hamiltonian and geometrical considerations.

\subsection{ Mean-curvature mass }

The expressions for the boundary Hamiltonians
\eqrefs{nthmcurvbt}{nthnullmcurvbt} in theorems~3.3 and~3.7,
respectively for timelike and null Hamiltonian flows, 
provide a one-parameter family of geometric definitions of
quasilocal mass-energy for regular 2-surfaces, 
\ie/ such that $S$ is closed, spacelike, and its mean curvature vector is 
spacelike or null, $\mcurv^2 \geq 0$. 

Some choice for the reference mass-energy density 
$\refHdens{\perp}(\sigma)$ 
must first be made in these expressions. 
The choice used in the mean-curvature mass expression \eqref{mcurvqlm} 
for the case $n=0$ 
has the obvious generalization
$\refHdens{\perp}(\sigma)= \abseuclmcurv^{n+1}= \abseuclP^{n+1}$
for $n>0$, 
under an isometric embedding of $(S,\sigma)$ 
into a spacelike hyperplane in Minkowski space, 
using $\twist\embed =0$ 
since the timelike normal in the embedding is 
parallel-transported under displacements
on the hyperplane. 

For spacelike 2-surfaces $S$ that are regular 
and have positive Gaussian curvature $\scurvS>0$, 
the {\it $n+1$-power mean-curvature mass} will be defined by 
\EQ\label{nthmcurvmass}
\qlE{\rm MC}_{(n)}(S) 
= \frac{c_0}{8\pi(n+1)} \sqrt{A}^n 
\oint_{S} (\abseuclmcurv^{n+1} - \absmcurv^{n+1}) \d{S} 
\doneEQ
(where the curvature positivity implies these 2-surfaces $S$ 
necessarily are restricted to have the topology of a 2-sphere). 

Another natural choice, 
extending the expression for the Hawking mass \eqref{hawkingqlm},
is $\refHdens{\perp}(\sigma)= \sqrt{4|\scurvS|}^{n+1}$,
which will apply to 2-surfaces $S$ of any topology. 

For spacelike 2-surfaces $S$ that are regular,
the {\it $n+1$-power Hawking mass} will be defined by 
\EQ\label{nthhawkingmass}
\qlE{H}_{(n)}(S) 
= \frac{c_0}{8\pi(n+1)} \sqrt{A}^n 
\oint_{S} (\sqrt{4|\scurvS|}^{n+1} - \absmcurv^{n+1}) \d{S} . 
\doneEQ

Both of the quasilocal mass expressions are well-defined for $n<-1$ 
only if $S$ is convex-regular, so $\absmcurv >0$ on $S$. 
However, for $n\geq 1$, these expressions can be applied
to marginally trapped surfaces or apparent horizons,
where $\absmcurv=0$ holds everywhere on $S$. 

Based on the boundary Hamiltonian in theorem~3.5,
the mass-energy definitions \eqrefs{nthmcurvmass}{nthhawkingmass}
can be generalized 
for convex-regular 2-surfaces further satisfying the local energy condition
$\absmcurv \geq |\twist|$:
\EQs 
\qlP{MC}_{(n)}(S) &&
= \frac{c_0}{8\pi (n+1)} \sqrt{A}^n 
\oint_{S} (\abseuclmcurv^{n+1} - \absP^{n+1}) \d{S} 
\nonumber\\&&
= \frac{c_0}{8\pi (n+1)} \sqrt{A}^n 
\oint_{S} (\abseuclmcurv^{n+1} - \sqrt{ \absmcurv^2 - |\twist|^2 }^{n+1}) \d{S}
\label{nthPmcurvmass}
\doneEQs
and 
\EQs
\qlP{H}_{(n)}(S) &&
= \frac{c_0}{8\pi (n+1)} \sqrt{A}^n 
\oint_{S} (\sqrt{4|\scurvS|}^{n+1} - \absP^{n+1}) \d{S} 
\nonumber\\&&
= \frac{c_0}{8\pi (n+1)} \sqrt{A}^n 
\oint_{S} (\sqrt{4|\scurvS|}^{n+1} 
- \sqrt{ \absmcurv^2 -|\twist|^2 }^{n+1}) \d{S} ,
\label{nthPhawkingmass}
\doneEQs
which will be called a symplectic variant of the quasilocal mass. 
Again for $n<-1$,
these expressions require that $S$ be sufficiently convex, 
so that $\absmcurv > |\twist| \geq 0$. 

Proposition~4.3:
The symplectic mass expressions satisfy 
\EQ
\qlP{}_{(n)}(S) \geq \qlE{}_{(n)}(S)
\doneEQ
with equality iff the 2-surface $S$ is twist-free, 
\ie/ $\twist=0$ everywhere on $S$. 

It is straightforward to fix $c_0$ in all these definitions 
by a comparison with the ADM mass in a large sphere limit,
as will be discussed later. 

The reference terms 
in the $n+1$-power Hawking and mean-curvature mass definitions 
\EQ\label{refterms}
\oint_{S} \sqrt{4|\scurvS|}^{n+1} \d{S} 
\quad\eqtext{ and }\quad
\oint_{S} |\mcurv\embed^{n+1} \d{S} 
\doneEQ
have a close relationship 
coming from the Gauss equation for spacelike 2-surfaces in spacetime
\EQ
2\scurvS = 
\frac{1}{2} \trk(\e{1}{})^2 - \frac{1}{2} \trk(\e{0}{})^2 
- |\shear{}{}(\e{1}{})|^2 + |\shear{}{}(\e{0}{})|^2
\doneEQ
where $|\shear{}{}(e)|^2 \equiv \shear{}{}(e)\cdot\shear{}{}(e)$. 
In a mean curvature frame for a regular 2-surface
embedded in a hyperplane in Minkowski space, 
the Gauss equation reduces to 
\EQ\label{gausseq}
\abseuclmcurv^2 = 
2\scurvS + 4|\shear{}{}(\unitmcurv)\embed^2 . 
\doneEQ
This is equivalent to the familiar Gauss equation 
in Euclidean surface theory \cite{ONeillII}. 
Consequently, 
if $S$ has positive curvature $\scurvS >0$
as assumed for existence of an embedding, 
the reference terms then differ only by a squared mean curvature shear term
$|\shear{}{}(\unitmcurv)\embed^2$
which vanishes when, and only when, $S$ is everywhere umbilic
in the hyperplane. 
Such umbilical surfaces $S$ can only be metric 2-spheres
(\ie/ isometric to a standard sphere in $\Rnum^3$ 
of radius $1/\sqrt{\scurvS}$)
as is well-known by the classification of 
compact umbilical 2-surfaces in Euclidean space \cite{ONeillII}. 

Proposition~4.4:
For a regular 2-surface $S$ with positive curvature, 
the reference terms \eqref{refterms} are equal iff $S$ is isometric to 
a metric 2-sphere. 
If $S$ is not a metric 2-sphere,
a comparison of the two reference terms yields the strict inequality 
\EQ
\qlE{MC}_{(n)}(S) = 
\frac{c_0}{8\pi(n+1)} \sqrt{A}^n 
\oint_{S} (\sqrt{4\scurvS + 2|\shear{}{}(\unitmcurv)\embed^2}^{n+1}
- \absmcurv^{n+1}) \d{S} 
> \qlE{H}_{(n)}(S)
\doneEQ
and likewise for the corresponding Dirichlet mass expressions. 

Clearly, 
the $n+1$-power mean-curvature mass \eqref{nthmcurvmass} 
will vanish for any 2-surface $S$ lying in a hyperplane in Minkowski space. 
In contrast, 
from Proposition~4.4 and the Gauss equation \eqref{gausseq}, 
the $n+1$-power Hawking mass \eqref{nthhawkingmass} 
vanishes only when such a 2-surface is umbilical. 
Namely, for non-umbilical 2-surfaces $S'$ that lie in a hyperplane, 
\EQ
\qlP{H}_{(n)}(S')
= \qlE{H}_{(n)}(S') 
= \frac{c_0}{8\pi(n+1)} \sqrt{A}^n 
\oint_{S'} ( \sqrt{ \abseuclmcurv^2 - 2 |\shear{}{}(\unitmcurv)\embed^2 }^{n+1}
- |\mcurv\embed^{n+1} )\d{S} 
< 0 
\doneEQ
as $\shear{}{}(\unitmcurv)\embed \neq 0$ 
on some portion of such a 2-surface $S'$. 
Notice this inequality holds in particular for the case of
the Hawking mass itself ($n=1$), 
\EQ
\qlE{H}(S') 
= -\frac{\sqrt{A}}{16\pi} \oint_{S'} |\shear{}{}(\unitmcurv)\embed^2 \d{S} 
< 0 . 
\doneEQ

In Minkowski space, 
hyperplanes are distinguished as hypersurfaces on which 
the dual mean curvature vector $\perpmcurv$ is parallel-transported
under displacements over the hypersurface,
\ie/ the mean curvature twist is identically zero on a hyperplane
and, in particular, on any 2-surface spanned by a hyperplane. 
A generalization of the latter property to any spacetime is provided by 
the definition of a twist-free 2-surface,
for which $\twist=0$ holds at every point.

\subsection{ Mean-curvature angular momentum }

The significance of a non-vanishing mean curvature twist $\twist$ 
for 2-surfaces $S$ 
is evidently connected with the Brown-York definition of 
quasilocal angular momentum 
when $S$ possesses a rotational Killing vector. 
A related meaning arises from 
the family of boundary Hamiltonians from theorem~3.6
in which $\twist$ is viewed as a geometric rotational flow 
on spacelike 2-surfaces.

For convex-regular 2-surfaces $S$, 
the {\it mean curvature angular momentum}
will be defined by 
\EQ\label{mcurvanglmom}
\qlJ{MC}(S)^2 
= {\tilde c_0} A^2 \oint_{S} |\twist|^2 \d{S} . 
\doneEQ
As with the mean-curvature mass, 
it is straightforward to fix $\tilde c_0$
by a comparison with the ADM angular momentum in a large sphere limit. 

The definition here involves the natural choice 
$\refHdens{\parallel}(\sigma) \equiv 0$
for the reference angular momentum density in the boundary Hamiltonian. 
This is motivated in terms of the rotational flow $\twist$
on $S$ by putting $\refHdens{\parallel}(\sigma)=|\twist\embed^{2}$ 
where $|\twist\embed=0$ 
under isometrically embedding $(S,\sigma)$ 
into a hyperplane in Minkowski space, 
paralleling what was done for the reference mass-energy density. 

The convexity condition on $S$, namely $\absmcurv>0$,
is necessary so a mean curvature normal frame exists
as needed to define $\twist$. 
There is a natural way to relax this condition
if $\twist$ is replaced by the null mean curvature twist $\twist^\pm$
which is a well-defined expression 
\eqrefs{nullmcurvtwistout}{nullmcurvtwistin}
in terms of the null mean curvature vectors 
$\mcurv_\pm$ for any regular spacelike 2-surface.
The boundary Hamiltonian in theorem~3.8 
motivates the following generalization of 
the mean curvature angular momentum.

For spacelike 2-surfaces that are regular,
the {\it null mean curvature angular momentum}
will be defined by 
\EQ\label{nullmcurvanglmom}
\qlJ{MC}_\pm(S)^2 
= {\tilde c_0} A^2 \oint_{S} |\twist^\pm|^2 \d{S} 
\doneEQ
which applies in particular when $S$ is a marginally trapped surface 
or an apparent horizon. 

The quasilocal quantity $\qlJ{MC}_\pm(S)^2$ 
should be compared with 
the angular momentum expression introduced by Hayward \cite{Hayward}, 
which differs by instead using the twist of a null frame fixed for $\TperpS$ 
in a certain manner chosen with respect to 
a 2+2 foliation of spacetime in terms of null hypersurfaces through $S$.
Changing this foliation produces a change in the Hayward's angular momentum,
while in contrast the angular momentum expression $\qlJ{MC}_\pm(S)^2$ 
depends only on $S$ and its extrinsic null geometry in spacetime. 

Finally, 
a slight generalization of these mean curvature angular momentum expressions
naturally arises from the scaled twists 
$\absmcurv^n\twist$ and $\absmcurv^n\twist^\pm$ 
interpreted as rotational flows on the 2-surface $S$:
\EQ\label{nthmcurvanglmom}
\qlJ{MC}_{(n)}(S)^2
= \tilde c_0 \sqrt{A}^{n+4} 
\oint_{S} \absmcurv^n |\twist|^2 \d{S} 
\doneEQ
and 
\EQ\label{nthnullmcurvanglmom}
\qlJ{MC}_{\pm(n)}(S)^2
= \tilde c_0 \sqrt{A}^{n+4} 
\oint_{S} \absmcurv^n |\twist^\pm|^2 \d{S} .
\doneEQ
Note these expressions respectively reduce to 
$\qlJ{MC}(S)^2$ and $\qlJ{MC}_\pm(S)^2$ when $n=0$
and are well defined for $n<0$ whenever $S$ is convex-regular 
(so $\absmcurv>0$ on $S$). 

For 2-surfaces $S$ that are merely regular (\ie/ non-convex),
$\qlJ{MC}_{(n)}(S)$ is well-defined for $n\geq 2$
due to the relation 
$\absmcurv^n\twist = \absmcurv^{n-2}\twist(\mcurv)$
where $\twist(\mcurv)$ is the twist of the mean curvature vectors 
$\perpmcurv,\mcurv$. 
Hence, $\qlJ{MC}_{(n)}(S)$ and $\qlJ{MC}_{\pm(n)}(S)$
are applicable respectively for $n\geq 2$ and $n\geq 0$
when $S$ is a marginally trapped surface or an apparent horizon.

\section{ Properties of mean-curvature mass and\\ angular momentum }
\label{properties}

Several criteria for a good definition of quasilocal mass
have been proposed by Christodoulou and Yau \cite{criteria}. 
Probably the most essential of their criteria, 
which should apply equally well to quasilocal angular momentum, 
are the following:
(i) agreement with ADM values in a large sphere limit at spatial infinity
and with irreducible mass at apparent horizons; 
(ii) zero for spacelike 2-surfaces in flat spacetime;
(iii) positivity under suitable conditions in curved spacetimes. 
These three properties will be addressed next 
for the family of mean-curvature mass and angular momentum definitions 
in \secref{quasilocal}.

\subsection{ Positivity }

An impressive positivity proof of
the mean-curvature mass \eqref {mcurvqlm} 
has been given recently by Liu and Yau \cite{LiuYau}
based on the Riemannian mean curvature theorem of Shi and Tam \cite{ShiTam}. 
Here a positivity result will be established for the family of 
mean curvature quasilocal masses \eqref{nthmcurvmass}: 
specifically, 
the higher-power mean-curvature masses ($n>0$) 
will be shown to be bounded below
due to positivity of the mean-curvature mass ($n=0$) itself,
and an alternative proof of the latter positivity will be given as well,
related to the spacelike mean curvature flow of 2-surfaces $S$ in spacetime.
In these results, $n$ will be restricted to non-negative integer values
for simplicity. 

Let $\Sigma$ be a spacelike hypersurface spanning a regular 2-surface $S$
with positive scalar curvature $\scurvS >0$ 
and hence with spherical topology. 
Throughout, it will be assumed that the spacetime curvature satisfies
the weak energy condition on $\Sigma$, 
so $\ricci(\n,\n) \geq -\frac{1}{2} \scurv$
where $\n$ is the future unit normal to $\Sigma$. 
In particular, the curvature inequality then holds
\EQ\label{weakenergy}
\R{}{} \geq \K{ij}{}\K{}{ij} -\K{i}{i}\K{j}{j}
\doneEQ
where, in coordinates adapted to $\Sigma$, 
$\R{}{}$ is the scalar curvature of $\h{ij}{}$, 
and $\K{ij}{} =\frac{1}{2}\Lie{\n}\h{ij}{}$ is the extrinsic curvature tensor. 
Let $\{\e{0}{},\e{1}{}\}$ be an oriented normal frame for $\TperpS$,
extended smoothly into $\Sigma$ such that $\e{0}{}$ is normal to $\Sigma$. 
Then the results of Shi and Tam show that 
the Brown-York quasilocal energy of $S$ is positive under the assumption
that the scalar curvature $\R{}{}$ of $\Sigma$ is nonnegative. 

Theorem~5.1 (Shi and Tam\cite{ShiTam}): 
If $\R{}{} \geq 0$ on $(\Sigma,\h{}{})$, then 
\EQ
\qlE{BY}(S,\e{0}{}) 
= \frac{1}{8\pi} \oint_{S} \trk(\e{1}{})\eucl - \trk(\e{1}{}) \d{S}
\geq 0
\doneEQ
with equality only when $(\Sigma,\h{}{})$ is isometric to 
Euclidean space $\Rnum^3$,
where $\trk(\e{1}{})\eucl$ is the extrinsic scalar curvature of $S$
under embedding into $\Rnum^3$ (by Weyl's theorem). 

Liu and Yau reduce the proof of positivity of the mean-curvature mass $(n=0)$
to theorem~5.1 on a fixed $\Sigma$ through constructing 
a conformal deformation of the hypersurface metric $\h{ij}{}$ by 
a procedure involving Jang's equation \cite{LiuYau}.
Here a different proof will be outlined by a special choice of 
the hypersurface $\Sigma$ motivated by properties of 
the dual mean curvature vector of $S$. 

At $S$, the extrinsic curvature tensor of $\Sigma$ decomposes into
\EQ
\K{ij}{} 
= -(\metric{i}{k} +\inve{1}{i} \e{1}{k}) (\metric{j}{l} +\inve{1}{j} \e{1}{l})
\D{k} \e{l}{0}
= \k{ij}{}(\e{0}{}) +\inve{1}{i} \perpwvec{}{j}(e)
+ 2\inve{1}{(i} \wvec{}{j)}(e)
\doneEQ
and hence
\EQ
\K{i}{i}
= -(\metric{}{ij} +\e{1}{i} \e{1}{j}) \D{i} \e{j}{0}
= \trk(\e{0}{}) + \perpwvec{}{1}(e) . 
\doneEQ
Consequently, 
observe that if $\Sigma$ is normal to $\perpmcurv$ at $S$ then 
the property $\trk(\unitperpmcurv)=0$ leads to 
\EQ
( \K{ij}{}\K{}{ij} -\K{i}{i}\K{j}{j} )|_S
= \kperp{ij}{} \kperp{}{ij} 
+ 2 \wvec{}{j} \wvec{j}{}
\geq 0
\doneEQ
with the notation
\EQ
\kperp{ij}{} \equiv \k{ij}{}(\unitperpmcurv) 
\doneEQ
for the (trace free) extrinsic curvature of $S$ 
in the dual mean curvature direction. 
This immediately implies, 
by the spacetime weak energy condition \eqref{weakenergy}, 
that 
\EQ
\R{}{}|_S \geq 0 . 
\doneEQ

Now suppose a hypersurface $\Sigma$ possesses a foliation
by a radial family of regular 2-surfaces $S(r)$ diffeomorphic to $S$
whose mean curvature vectors $\mcurv(r)$ lie in $\Sigma$. 
(The 2-surfaces $S(r)$ are assumed to have 
monotonically decreasing area $A(r)=\int_{S(r)} \d{S}$, 
parametrized by the radial coordinate
$r=\sqrt{A(r)/(4\pi)}$, with the limit $S(0)$ as $r\rightarrow 0$
denoting a center point for the foliation, \ie/ 
$\Sigma \simeq S(0) \cup S\times \Rnum$.)
Such a hypersurface will be called {\it sectionally-maximal} since 
the dual mean curvature vectors $\perpmcurv(r)$ will define 
a hypersurface orthogonal vector field 
and hence each 2-surface $S(r)$ will have zero intrinsic expansion 
orthogonal to $\Sigma$ in spacetime, $\trkperp|_{S(r)} =0$. 

Lemma~5.2:
On a sectionally-maximal hypersurface $\Sigma$,
the scalar curvature is nonnegative
\EQ
\R{}{} \geq
\kperp{ij}{} \kperp{}{ij} + 2 \wvec{}{j} \wvec{j}{}
\geq 0
\doneEQ
whenever the weak energy condition holds on the spacetime curvature. 
Moreover, the mean curvature of $S$ in $\Sigma$ equals
the spacetime mean curvature of $S$, $\trk(\unitmcurv) = \absmcurv$. 

Combining this result with the theorem of Shi and Tam
yields positivity of the mean-curvature mass. 

Theorem~5.3:
Let $S$ be a regular 2-surface spanned by 
a sectionally-maximal spacelike hypersurface $\Sigma$
on which the weak energy condition is satisfied. 
Then the mean-curvature mass of $S$ is nonnegative
\EQ
\qlE{MC}(S) = 
\frac{1}{8\pi} \oint_{S} \abseuclmcurv - \absmcurv \d{S} 
\geq 0
\doneEQ
with equality only when $(\Sigma,\h{}{})$ is isometric to 
Euclidean space $\Rnum^3$. 

Given any regular 2-surface $S$ in spacetime,
a geometrically natural means for addressing existence of 
a sectionally-maximal spacelike hypersurface $\Sigma$ 
is to consider the inward spacelike mean curvature flow of $S$.
The flow is defined by Lie dragging $S$ along its inward
mean curvature vector field $\mcurv$ in spacetime
\EQ
\Lie{r} \x{\mu}{}(S(r)) \propto - \Hvec{\mu}{}(r)
\doneEQ
where $\x{\mu}{}(\cdot)$ denotes the embedding of $S$ in coordinates on $M$.
This is a generalization of the Riemannian mean curvature flow
studied by Huisken \cite{Huisken,mcurvflow},
where the flow 
$\Lie{r} \x{i}{}(S(r)) \propto - \Hvec{i}{}(r) = \trk(\u) \uvec{i}{}$
takes place in a fixed 3-dimensional Riemannian manifold $(\Sigma,\h{}{})$
and $\u$ is the inward unit normal of $S(r)$ in $\Sigma$. 
Under suitable geometric assumptions on $(S,\sigma)$ and $(\Sigma,\h{}{})$
the Riemannian mean curvature flow $S(r)$ exists and is smooth,
with $S(r)$ contracting to a center point $S(0)$ in a finite time
(such that $S(r)$ becomes diffeomorphic to 
a metric sphere of area $4\pi r^2$ as $r\rightarrow 0$). 
A similar result might be expected to hold for 
the mean curvature flow of $S$ in spacetime $(M,g)$
under sufficiently strong geometric conditions on $(S,\sigma)$ and $(M,g)$.
A key issue will be that 
the flow must be shown to remain spacelike and smooth 
sufficiently long to reach a center point $S(0)$, 
\ie/ $\Hvec{\mu}{}\Hvec{}{\mu} >0$ for all $r>0$. 
(Outward spacelike mean curvature flow of 2-surfaces $S$ in
asymptotically flat spacetimes has been considered recently in \Ref{mcurvflow}
in connection with the Penrose inequality.)

Positivity of mean-curvature mass in theorem~5.3 
leads to a (negative) lower bound
on the higher-power mean-curvature masses. 
As motivation for the nature of the bound, 
note the sign of the mass density 
$\abseuclmcurv^{n+1} - \absmcurv^{n+1}$ on a 2-surface $S$ 
is controlled by 
the difference in scalar mean curvatures $\abseuclmcurv$ and $\absmcurv$.
Let $S_\pm$ denote, respectively, the portions
(closed, measurable subsets) of $S$
on which $\pm\abseuclmcurv \geq \pm\absmcurv$, 
so $\qlE{MC}(S) = \qlabsE{}(S_+)- \qlabsE{}(S_-) \geq 0$
where 
$\qlabsE{}(S_\pm) \equiv 
\frac{1}{8\pi} \int_{S_\pm} | \abseuclmcurv - \absmcurv | \d{S} \geq 0$.
The aim will be to use the preceding positivity inequalities 
to give a lower bound on $\qlE{MC}_{(n)}(S)$
proportional to $\qlabsE{}(S_-)$ for any $n$, 
since $\qlE{MC}_{(n)}(S)$ is manifestly non-negative if $S_-$ is empty. 

Theorem~5.4:
If $S$ satisfies the hypothesis in theorem~5.3, 
then the higher-power mean-curvature mass of $S$ is bounded below by 
\EQ
(i)\qquad
\qlE{MC}_{(n)}(S) 
\geq -((\max\absmcurv)^{n} -(\min\absmcurv)^{n}) 
c_0 \sqrt{A}^n \qlabsE{}(S_-) ,
\quad
n\geq 0
\doneEQ
where $\max\absmcurv=\sup_S(\absmcurv)$, $\min\absmcurv=\inf_S(\absmcurv)$,
which are finite since $S$ is closed. 
For $n\geq 1$ a sharper bound is given by 
\EQs
(ii)\qquad
\qlE{MC}_{(n)}(S)/(c_0 \sqrt{A}^n) 
&& \geq 
| ((8\pi/A)\qlE{MC}(S) +\min\absmcurv)^{n+1} - (\min\absmcurv)^{n+1} |
A/8\pi(n+1)
\nonumber\\&&\qquad
- | (c+1)(\max\absmcurv)^{n} -c(\min\abseuclmcurv)^{n} | \qlabsE{}(S_-) 
\doneEQs
for some positive constant $c$ depending only on $n$. 

Proof. 
Write $\mcurv_0 = \abseuclmcurv$, 
$\minH= \min\absmcurv$, $\maxH= \max\absmcurv$. 
Note the positivity $\qlE{MC}(S) \geq 0$
implies the inequality 
\EQ
\int_{S_+} \mcurv_0 - \absmcurv \d{S} 
\geq 
\int_{S_-} \absmcurv - \mcurv_0 \d{S} . 
\label{Hineq}
\doneEQ
In addition, note the pointwise inequalities
\EQ
\mcurv_0 \geq \minH \geq 0 \eqtext{ on $S_+$ } ,\quad
0\leq \mcurv_0 \leq \maxH \eqtext{ on $S_-$ } .
\label{pointHineq}
\doneEQ
So
\EQs
\int_S \mcurv_0^\pow -\absmcurv^\pow \d{S}
&&
= \int_S (\mcurv_0 -\absmcurv) \sum_{k=1}^\pow 
\mcurv_0^{\pow-k} \absmcurv^{k-1} \d{S}
\nonumber\\
&&
\geq 
\pow \minH^{\pow-1} \int_{S_+} \mcurv_0 -\absmcurv \d{S}
- \pow \maxH^{\pow-1} \int_{S_-} \absmcurv -\mcurv_0 \d{S} . 
\label{Hnineq}
\doneEQs
Then we combine inequalities \eqrefs{Hineq}{Hnineq} and put $\pow=n+1$,
which establishes (i) for $n\geq 0$. 
To derive the sharper result (ii), 
note 
\EQ
\int_S \mcurv_0^\pow -\absmcurv^\pow \d{S}
= \sum_{k=1}^\pow {\pow\choose k} 
\int_S \absmcurv^{\pow-k} (\mcurv_0 -\absmcurv)^k \d{S} . 
\label{Hnid}
\doneEQ
The terms in integral \eqref{Hnid} with even powers of $\mcurv_0 -\absmcurv$
are manifestly positive and can be bounded below by means of 
the Holder inequality
\EQ
8\pi \qlE{MC}(S)
\leq \int_S |\mcurv_0 -\absmcurv| \d{S}
\leq A^{1-1/k} (\int_S (\mcurv_0 -\absmcurv)^k \d{S})^{1/k}
\quad \eqtext{for $k$ even}
\doneEQ
combined with the elementary inequality 
\EQ
\int_S \absmcurv^k f \d{S}
\geq \minH^k \int_S f \d{S}
\doneEQ
holding for any nonnegative function $f$ on $S$. 
A bound for the terms with odd powers of $\mcurv_0 -\absmcurv$
in integral \eqref{Hnid} is obtained by means of 
the positivity inequalities
\EQ
\int_{S_+} \mcurv_0 - \absmcurv \d{S} 
\geq 
\int_{S} \mcurv_0 - \absmcurv \d{S} 
\geq 0 ,\quad
A \geq A(S_+)
\label{Hpos}
\doneEQ
as follows. 
We split the integral into portions over $S_\pm$. 
By steps similar to those before, for all odd $k$, 
\EQs
\int_{S_+} \absmcurv^{\pow-k} (\mcurv_0 -\absmcurv)^k \d{S}
&& 
\geq 
\minH^{\pow-k} \int_S |\mcurv_0 -\absmcurv|^k \d{S}
\nonumber\\
&&
\geq 
\minH^{\pow-k} A(S_+)^{1-k} (8\pi \qlE{MC}(S_+))^k
\nonumber\\
&&
\geq 
\minH^{\pow-k} A^{1-k} (8\pi \qlE{MC}(S))^k . 
\doneEQs
Next, 
\EQs
\int_{S_-} \absmcurv^{\pow-k} (\mcurv_0 -\absmcurv)^k \d{S}
&& 
= -\int_{S_-} \absmcurv^{\pow-k} |\mcurv_0 -\absmcurv|^{k-1} 
(\mcurv_0 -\absmcurv) \d{S}
\nonumber\\
&&
\geq 
-\maxH^{\pow-k} 2^{k-2} (\maxH^{k-1} -\minH{}^{k-1}\embed) 
|8\pi\qlE{MC}(S_-)|
\nonumber\\
&&
\geq 
-2^{k-2} (\maxH^{\pow-1} -\minH{}^{\pow-1}\embed) 
|8\pi\qlE{MC}(S_-)|
\doneEQs
as we see from the pointwise inequality \eqref{pointHineq}
combined with the algebraic inequality
\EQ
(\absmcurv -\mcurv_0)^{k-1} 
\leq  
2^{k-2} (\absmcurv^{k-1} -\mcurv_0^{k-1}) 
\doneEQ
which holds for $\absmcurv \geq \mcurv_0 >0$ and $k>1$. 
Similarly, for $k=1$, 
\EQ
\int_{S_-} \absmcurv^{\pow-1} (\mcurv_0 -\absmcurv) \d{S}
\geq 
-\maxH^{\pow-1} |8\pi\qlE{MC}(S_-)| . 
\doneEQ
Adding the previous inequalities, we obtain (ii). 

These positivity results in theorems~5.3 and~5.4 
extend to the quasilocal Dirichlet masses. 

Proposition~5.5:
If the local energy condition 
$\absmcurv \geq |\twist|$
holds on a regular 2-surface $S$, 
then 
\EQ
\qlP{MC}_{(n)}(S) 
\geq \qlE{MC}_{(n)}(S) . 
\doneEQ

\subsection{ Large sphere limit and apparent horizons }

The mean-curvature mass and angular momentum have good behavior 
in the limit $S_\infty$ of spherical regular 2-surfaces $S$ 
that suitably approach spatial infinity in asymptotically flat spacetimes. 
This allows the normalization constants $c_0$ and $\tilde c_0$ 
to be chosen so the definitions match the corresponding ADM quantities 
at spatial infinity. 

Calculation of the large sphere limit follows straightforwardly
from results in \Ref{paperII} for
the mean curvature $\absmcurv$ and twist $\twist(\mcurv)$
in the large sphere limit. 

Proposition~5.6:
Let $M_\infty$, $J_\infty$ denote the ADM mass and angular momentum
for an asymptotically flat spacetime \cite{ChoquetBruhat}. 
Then the $n+1$-power mean-curvature mass satisfies
\EQ
\qlE{MC}_{(n)}(S_\infty) 
= M_\infty
\doneEQ
if 
$c_0= 1/\sqrt{16\pi}^n$, 
while the $n+1$-power mean-curvature angular momenta satisfy
\EQ
\qlJ{MC}_{(n)}(S_\infty) 
= \qlJ{MC}_{\pm(n)}(S_\infty) 
= J_\infty
\doneEQ
if 
$\tilde c_0= 32/(3\sqrt{16\pi}^{n+6})$.

The mean-curvature mass and angular momentum 
also have satisfactory behavior at apparent horizons, 
namely when $S$ is a spacelike 2-surface on which $\absmcurv=0$.

Proposition~5.7:
For marginally trapped spacelike 2-surfaces $S_{\rm hor}$, 
the $n+1$-power mean-curvature mass is equal to 
\EQ
\qlE{\rm MC}_{(n)}(S_{\rm hor})
= \frac{c_0}{8\pi(n+1)} \sqrt{A}^n 
\oint_{S} \abseuclmcurv^{n+1} \d{S} >0 .
\doneEQ

A good lower bound for this mass arises from the Minkowski inequality
\EQ
\oint_{S} \abseuclmcurv \d{S} \ge \sqrt{16\pi A} = A/\mirr
\doneEQ
holding for any convex 2-surface $S$ embedded
in a hyperplane in flat spacetime, 
where $A$ is its area
and $\mirr =\sqrt{A/(16\pi)}$ is its irreducible mass
\cite{criteria}. 
Equality holds iff $S$ is a standard sphere. 
Holder's inequality then leads to an analogous bound on 
the $n+1$-power mean curvature integral 
\EQ
\frac{1}{A} \oint_{S} \abseuclmcurv^{n+1} \d{S} 
\ge
\left( \frac{1}{A} \oint_{S} \abseuclmcurv \d{S} \right)^{n+1}
\ge
\sqrt{16\pi/A}^{n+1} = 1/\mirr^{n+1} .
\doneEQ
This immediately implies a lower bound on 
the $n+1$-power mean-curvature mass
for any marginally trapped 2-surface,
\EQ\label{bound}
\qlE{\rm MC}_{(n)}(S_{\rm hor})
\ge
\frac{c_0}{8\pi(n+1)} \sqrt{16\pi}^n \sqrt{16\pi A}
=\frac{2\mirr}{n+1}
\doneEQ
where $c_0$ is normalized according to the large sphere limit 
in proposition~5.6.
In particular, 
the mean-curvature mass is bounded below precisely by
the irreducible mass when $n=1$,
so in this case it obeys a quasilocal version of the
well-known Gibbons-Penrose inequality. 
This inequality is slightly violated 
in the higher power cases $n>1$
by the factor $2/(n+1) <1$, 
\ie/ the lower bound \eqref{bound} is a weaker inequality. 
Interestingly, for $n=0$ 
there is also a violation of the Gibbons-Penrose inequality 
but with a factor $2$, 
namely the lower bound \eqref{bound} yields a stronger inequality. 
Nevertheless, in all cases the mean-curvature mass 
is bounded below in terms of a multiple of $\mirr$.

The behavior of mean-curvature angular momentum is less straightforward.
Recall, for any apparent horizon, 
the mean curvature vectors become degenerate and null,
$\mcurv=\pm\perpmcurv$ 
(respectively in the outward/inward trapped cases). 
Thus $\twist(\mcurv)=0$ vanishes on $S_{\rm hor}$,
since 
$v\hook\twist(\mcurv)=\mcurv\cdot\covder{v}\perpmcurv=
\pm 2\covder{v}\absmcurv=0$
for all tangent vectors $v$ on $S_{\rm hor}$. 
Consequently, 
the $n+1$-power mean-curvature angular momentum 
$\qlJ{MC}_{(n)}(S_{\rm hor})$
vanishes for $n\ge 2$,
which is precisely when it is well-defined for apparent horizons. 
In particular, it is not well-defined for $n=0$, 
since $\twist$ becomes singular whenever $\absmcurv=0$ on a 2-surface.
On the other hand, as $\twist^\pm$ is non-singular, 
the $n+1$-power null mean-curvature angular momentum 
$\qlJ{MC}_{\pm(n)}(S_{\rm hor})$ 
is well-defined and non-vanishing for $n=0$
(while it vanishes for $n>0$). 

More insight into the interpretation of these mean-curvature angular momenta
comes from considering horizons of stationary black holes,
related to Killing horizons \cite{Wald-book2}.

\subsection{ Killing horizons and static 2-surfaces }

Recall, a Killing vector is hypersurface orthogonal in spacetime iff
the corresponding covector 
(identifying $\TM$ and $\coTM$ via the metric $g$)
obeys the integrability condition \cite{Wald-book}
\EQ\label{orthokv}
\covder{}\kv = f\wedge\kv
\doneEQ
for some smooth 1-form $f$ in spacetime, 
where the symmetric part of $\covder{}\kv$ vanishes by
the Killing equation on $\kv$. 
It is useful to generalize this notion to consider 2-surfaces 
spanned by a local hypersurface that is orthogonal to
a Killing vector. 

Definition~5.8:
A Killing vector $\kv$ is said to be 2-surface orthogonal 
if, for a spacelike 2-surface $S$,
$\kv$ is normal to $S$ 
and satisfies the hypersurface-orthogonality condition \eqref{orthokv}
everywhere on $S$. 
Such a 2-surface will be called static with respect to $\kv$. 

As examples encompassing stationary black-hole horizons, 
spatial cross-sections of a Killing horizon \cite{Wald-book2} in spacetime 
produce spacelike 2-surfaces $S$ that are static with respect to 
the null Killing vector generating the horizon. 
For such situations where a 2-surface orthogonal Killing vector $\kv$
is null at $S$, $\kv^2|_S=0$, 
the twist of $\kv|_S$ is directly related to the null mean curvature twist
\EQ
\twist(\kv) = \twist^\pm +\covSder{}\ln|\alpha|
\doneEQ
where $\alpha$ is a smooth function on $S$ determined by 
$\kv|_S =\alpha\mcurv_\pm$
when the 2-surface $S$ is respectively outward/inward marginally trapped. 
Moreover, the 2-surface orthogonality condition shows that 
$\alpha$ can be expressed in terms of the surface gravity 
$\kappa_{\rm hor} \equiv \frac{1}{2}(\covder{\chi}\kv^2)|_S$
by 
\EQ
\alpha = 
\frac{1}{2\kappa_{\rm hor}} (\covder{\mp}\kv^2)|_S
\doneEQ
in the outward/inward trapped cases, 
where $\chi$ is a vector such that $(\kv\wedge\chi)|_S=\perpvol{}{}(S)$,
and where $\covder{\mp} =(\mcurv_\mp\cdot\covder{})|_S$.
Hence the null mean-curvature angular momentum 
$\qlJ{MC}_{\pm}(S)$
for static 2-surfaces with respect to a null Killing vector $\kv$
is naturally related to the surface gravity and twist of $\kv$ at $S$. 

Other physically relevant examples of static 2-surfaces are given by 
any 2-surface $S$ contained in a spacelike hypersurface in 
a static spacetime with a timelike Killing vector. 
In such situations where a 2-surface orthogonal Killing vector $\kv$
is timelike at $S$, $\kv^2|_S <0$, 
the twist of $\kv|_S$ vanishes on $S$ as follows from
the 2-surface orthogonality condition:
\EQ
v\hook\twist(\kv) 
= ({*\kv}\cdot\covder{v}\kv)|_S 
= ({*\kv}\cdot(v\hook(f\wedge\kv)))|_S 
=0
\doneEQ
for all tangential vectors $v$ on $S$,
since $v\cdot\kv|_S={*\kv}|_S\cdot\kv|_S=0$. 
Now, 
through the geometric relation $\kv|_S = \alpha \perpmcurv$
where $\alpha$ is a smooth non-vanishing function on $S$, 
the mean curvature twist is found to vanish,
\EQ
\twist = \alpha^{-2}\absmcurv^{-2} \twist(\kv) = 0 . 
\doneEQ
Then the null mean curvature twist reduces to a gradient,
$\twist^\pm = \mp\frac{1}{2}\covSder{}\ln\absmcurv$. 

Proposition~5.9:
For static 2-surfaces $S$ with respect to a timelike Killing vector $\kv$,
the $n+1$-power mean-curvature angular momentum 
$\qlJ{MC}_{(n)}(S)$ is zero,
while the $n+1$-power null mean-curvature angular momentum is given by 
\EQ
\qlJ{MC}_{\pm(n)}(S) = 
\frac{\tilde c_0}{4} \sqrt{A}^{n+4}
\oint_S \absmcurv^{n-2}(\covder{}\absmcurv)^2 \d{}S . 
\doneEQ
Hence $\qlJ{MC}_{\pm (n)}(S)$ is zero iff $S$ has CMC.

\subsection{ Flat spacetime }

Clearly, 
the mean-curvature mass and angular momenta
are identically zero for any 2-surface 
lying in a hyperplane in Minkowski space. 
This reflects the desirable physical property that
physical values of quasilocal mass and angular momentum 
assigned to spacelike 2-surfaces in Minkowski space should be zero,
because there is no gravitational field. 
But this good behavior of the mean-curvature mass and angular momentum 
is no longer true in general for 2-surfaces that lie 
in curved spacelike hypersurfaces in Minkowski space. 
In particular, 
\'OMurchadha, Szabados, Tod \cite{lightconeex}
have recently shown that spacelike cuts of a light cone in Minkowski space
yield positive curvature 2-surfaces some of which 
have nonzero mean-curvature mass. 
The mean-curvature angular momentum of such 2-surfaces also is nonzero,
since the twist along a cut given by a curved hypersurface 
does not identically vanish. 
(In fact, as is well known \cite{Brinkmann}, 
any spacelike 2-surface always admits, at least locally,
an isometric embedding into a light cone in Minkowski space.
For such an embedding the mean-curvature mass and angular momentum 
will, in general, be nonzero.)
Calculation of the pair of mean curvature vectors $\perpmcurv,\mcurv$,
and the twist $\twist$ 
for arbitrary spacelike cuts of a light cone in Minkowski space 
are given in \Ref{paperII}.

This situation is readily understood from the fact that, roughly speaking, 
a one-function degree of freedom characterizes the existence of 
a local isometric embedding of a given spacelike 2-surface 
into flat spacetime,
as seen by a simple function counting argument 
using the Gauss-Codacci equations. 
In particular, taking into account the coordinate freedom, 
this system of equations is underdetermined 
as there is one more unknown than the number of equations.

\section{ Concluding remarks }

On physical grounds, 
the expressions introduced 
for quaslocal mean-curvature masses \eqsref{nthmcurvmass}{nthPhawkingmass}
and angular momenta \eqsref{mcurvanglmom}{nthnullmcurvanglmom}
are not completely satisfactory because 
they assign nonzero values to some spacelike 2-surfaces
when no gravitational field is present in a region of spacetime 
containing the 2-surface. 
Nevertheless, 
these definitions have a natural motivation 
from geometric and Hamiltonian considerations
and so they may well be relevant at least in mathematical contexts e.g. 
in the study of the Cauchy problem for the Einstein field equations 
as discussed by Liu and Yau \cite{LiuYau};
or also in certain physical applications that 
are insensitive to the flat spacetime values 
as explained in \Ref{canonical}. 

Of course it is worthwhile to explore whether these definitions 
can be improved to have more physically satisfactory behavior
for spacelike 2-surfaces in curved hypersurfaces in flat spacetime. 
The most direct approach is to try 
imposing sufficient extra conditions under which a regular 2-surface 
will have a {\it rigid} isometric embedding into Minkowski space
(\ie/ unique up to isometry motions). 
In this case the embedding can then be used to adjust
the reference subtraction terms to make
the mean-curvature mass and angular momentum vanish
for all such 2-surfaces in flat spacetime. 

One natural proposal would be to use 
the local energy condition $\P^2\leq 0$ 
as a restriction on spacelike 2-surfaces. 
Note this condition $\absmcurv \geq |\twist|$ holds trivially 
for 2-surfaces $S$ that lie in any hyperplane in Minkowski space,
since $\twist=0$ on such $S$. 
So, suppose $S$ is a spacelike 2-surface in Minkowski space 
satisfying $\P^2 \leq 0$, 
such that the only isometric embeddings of it back into Minkowski space
that obey $\absembedmcurv \geq |\twist\curvembed$ 
are rigid (isometry) motions of $S$.
Then its adjusted mean-curvature mass would vanish,
and furthermore, 
its mean-curvature angular momentum could be adjusted to vanish 
by subtraction of a suitable reference density term 
involving $|\twist\curvembed$ given by 
\EQ\label{qlEadjust}
\qlE{MC'}_{(n)}(S)
= \frac{c_0}{8\pi (n+1)} \sqrt{A}^n 
\oint_{S} (\absembedmcurv^{n+1} - \absmcurv^{n+1}) \d{S} 
\doneEQ
and similarly 
\EQ\label{qlJadjust}
\qlJ{MC'}_{(n)}(S)^2
= {\tilde c_0} A^2 \oint_{S} (|\twist|^2 - |\twist\curvembed^2) \d{S}  
\doneEQ
where $\absembedmcurv$ and $|\twist\curvembed$ 
refer to the assumed embedding of $S$ in Minkowski space. 
If such a 2-surface $S$ is contained in a flat spacetime region 
then both the adjusted mean-curvature mass \eqref{qlEadjust} 
and angular momentum \eqref{qlJadjust} are zero. 

In this situation the symplectic variant of the mass also has good behavior
if its reference term is adjusted similarly
\EQs
\qlP{MC'}_{(n)}(S) &&
= \frac{c_0}{8\pi (n+1)} \sqrt{A}^n 
\oint_{S} (\absembedP^{n+1} - \absP^{n+1}) \d{S} 
\label{qlPadjust}\\&&
= \frac{c_0}{8\pi (n+1)} \sqrt{A}^n 
\oint_{S} 
( \sqrt{ \absembedmcurv^2 - |\twist\curvembed^2 }^{n+1} 
- \sqrt{ \absmcurv^2 - |\twist|^2 }^{n+1} )\d{S} . 
\nonumber
\doneEQs

A different proposal can be made, 
using just the twist of a 2-surface in spacetime to determine 
a reference subtraction by an isometric embedding of the 2-surface 
into Minkowski space 
without the need to restrict 2-surfaces a priori 
by any local energy condition. 
Namely, 
consider isometric embeddings of $(S,\sigma)$ into Minkowski space 
preserving the twist $\twist$ 
(and hence the normal curvature $\norscurvS\vol{}{}(S) =\d{}_S\twist$).  
This will be called a {\it twisted-embedding}. 
(It is worth noting that preservation only of $\norscurvS$ itself 
is a strictly weaker requirement, 
since the results in \Ref{paperII} show that 
for light cone 2-surfaces the twist is a gradient 
and thus $\norscurvS=0$ but $\twist \neq 0$;
such an embedding condition has been discussed previously in \Ref{Epp}.)
Although little seems to be known about 
the existence of embeddings of $(S,\sigma,\twist)$ in general, 
it is plausible to expect that with sufficient assumptions 
a regular 2-surface $S$ 
will admit a unique embedding of this kind (up to rigid motions of $S$). 
At the least, function counting shows that
with both $\sigma$ and $\twist$ specified,
the Gauss-Codacci equations are no longer an underdetermined system. 

In this setting 
the adjusted mean-curvature mass expressions \eqrefs{qlEadjust}{qlPadjust}
carry over 
with $\mcurv\curvembed$ now denoting 
the scalar mean curvature of $S$ in the twisted-embedding, 
and with $\twist\curvembed=\twist$. 
These expressions have a direct Hamiltonian interpretation 
if the symplectic vector $\P$ is included as part of the boundary data
for a Hamiltonian variational principle in theorems~3.3 and~3.5. 
In particular, 
this allows the reference term $\refHdens{\perp}(\sigma,\P)$ appearing 
in the Hamiltonians to depend on the twist $\twist=\P|_{\TS}$
of the 2-surface. 
(Some additional motivation for the idea of twisted-embeddings 
comes from a spinor formulation of the Hamiltonian and 
an analysis of its positivity \cite{spinoranalysis}.)

Lemma~6.1:
If a regular 2-surface $S$ admits 
a rigid (up to isometry) twisted-embedding into Minkowski space, 
then $\qlE{MC'}_{(n)}(S)=\qlP{MC'}_{(n)}(S)=0$
when spacetime is flat in any region containing $S$. 

Under a twisted-embedding of a 2-surface $S$,
the definitions \eqrefs{qlEadjust}{qlPadjust} will be called 
the ($n+1$-power) {\it twisted mean-curvature mass}.
Their behavior is obviously more satisfactory in giving 
a quasilocal mass of zero to all suitable spacelike 2-surfaces $S$ 
contained in any flat spacetime region
and they reduce to the ($n+1$-power) mean-curvature mass 
in the case of twist-free 2-surfaces. 
There is a trade off, however, concerning their positivity. 
Due to the adjustment of the reference subtraction terms,
positivity of mean-curvature mass in theorem~5.3 
gives rise only to an upper bound on the twisted mean-curvature mass
as follows. 

Proposition~6.2:
If $S$ is a regular 2-surface
with positive scalar curvature, 
then under any isometric embedding of $(S,\sigma)$ into Minkowski space, 
\EQ
\qlE{MC}(S) 
\geq \qlE{MC'}(S) . 
\doneEQ

Proof. 
The difference in masses reduces to 
\EQ
\qlE{MC}(S) - \qlE{MC'}(S) 
= \frac{1}{8\pi} \oint_S (\abseuclmcurv -\absembedmcurv) \d{S}
= \qlE{MC}(S\curvembed) \geq 0
\doneEQ
due to positivity of mean-curvature mass, 
where $S\curvembed$ refers to $S$ as embedded in Minkowski space. 

Regarding the mean-curvature angular momentum expressions, 
because $\twist$ is preserved, 
$\qlJ{MC}(S)$ for 2-surfaces $S$ in flat spacetime will be zero 
only in the special case of twist-free 2-surfaces, 
\ie/ when $\twist(\mcurv)=0$ on $S$.
Moreover, an adjusted expression, 
$\qlJ{MC'}_{(n)}(S)$ as above, 
would trivially vanish for all 2-surfaces $S$ 
possessing a twisted-embedding.
Consequently, in this situation 
$\qlJ{MC}(S)$ will no longer be connected with 
quasilocal angular momentum, 
but instead will become an invariant of the twisted-embedding.

\begin{acknowledgments}
Roh Tung is thanked for contributions and fruitful discussions 
in the course of this research,
particularly for an initial outline of the Noether charge analysis. 
The referee is thanked for valuable comments which have
improved the presentation and results. 
\end{acknowledgments}

\def\v#1{{\bf #1}}


\begin{thebibliography}{99}


\bibitem{paperI}
S.C. Anco and R.S. Tung,
``Covariant Hamiltonian boundary conditions in General Relativity 
for spatially bounded spacetime regions'',
J. Math. Phys. \v{43}, (2002) 5531--5566.

\bibitem{paperII}
S.C. Anco and R.S. Tung,
``Properties of the symplectic structure of general relativity 
for spatially bounded spacetime regions'',
J. Math. Phys. \v{43}, (2002) 3984--4019. 

\bibitem{errata}
errata, 
J. Math. Phys. \v{45}, (2004) 2108--2109.

\bibitem{bterms}
J.W. York, 
``Boundary terms in the Action Principles of General Relativity'',
Found. Phys. \v{16}, (1986) 249--257.

\bibitem{BrownYork1}
J.D. Brown and J.W. York,
``Quasilocal energy in General Relativity''
in {\it Mathematical aspects of classical field theory}, 
Contemp. Math. (AMS) \v{132}, (1992) 129--142. 

\bibitem{BrownYork2}
J.D. Brown and J.W. York,
``Quasilocal energy and conserved charges derived from the 
gravitational action'', 
Phys. Rev. D \v{47}, (1993) 1407--1419. 

\bibitem{canonical}
J.D. Brown, S.R. Lau, J.W. York,
``Action and Energy of the Gravitational Field'',
Annals Phys. \v{297} (2002) 175--218.

\bibitem{Nester1}
C.M. Chen, J.M. Nester, R.S. Tung, 
``Quasilocal energy-momentum for geometric gravity theories'',
Phys. Lett. A \v{203} (1995), 5--11; 
C.M. Chen and J.M. Nester, 
``Quasilocal quantities for GR and other theories of gravity'',
Class. Quant. Grav. \v{16} (1999), 1279--1304. 

\bibitem{ONeillI}
B. O'Neill,
{\it Semi-Riemannian Geometry},
(Academic Press: New York) (1983).

\bibitem{Chen}
B.-Y. Chen,
{\it Geometry of submanifolds},
(Dekker: New York) (1973).

\bibitem{Kijowski}
J. Kijowski, ``A Simple Derivation of Canonical Structure and Quasi-Local
Hamiltonians in General Relativity'',
Gen. Rel. Grav. \v{29}, (1997) 307--343;  
J. Kijowski and W.M. Tulczyjew,
{\it A Symplectic Framework for Field Theories},
Lecture Notes in Physics No. 107
(Springer-Verlag: Berlin) (1979).

\bibitem{Lau}
S.R. Lau, 
``New variables, the gravitational action, and boosted quasilocal 
stress-energy-momentum'', 
Class.Quant.Grav. \v{13}, (1996) 1509--1540. 

\bibitem{Epp}
R.J. Epp,
``Angular momentum and an invariant quasi-local energy in General
Relativity",
Phys. Rev. D62 (2000) 124018.

\bibitem{LiuYau}
C.-C. Liu and S.-T. Yau,
``Positivity of quasilocal mass'',
Phys. Rev. Lett. \v{90}, (2003) 231102.

\bibitem{ShiTam}
Y. Shi and L.-F. Tam, 
``Positive mass theorem and the boundary behaviors of
compact manifolds with nonnegative scalar curvature'',
J. Diff. Geom. \v{62}, (2002) 79--125.

\bibitem{Wald-book}
R.M. Wald,
{\it General Relativity}
(University of Chicago Press: Chicago) (1984).

\bibitem{HawkingEllis}
S.W. Hawking and G.F.R. Ellis, 
{\it The large scale structure of space-time}
(Cambridge University Press: Cambridge) (1973).

\bibitem{horizons}
A. Ashtekar and B. Krishnan,
``Dynamical horizons: energy, angular momentum, fluxes, and balance laws'',
Phys. Rev. Lett. \v{89}, (2002) 261101;
A. Ashtekar and C. Beetle,
``Mechanics of rotating isolated horizons'',
Phys. Rev. D \v{64}, (2001) 044016;
A. Ashtekar, S. Fairhurst, B. Krishnan,
``Isolated horizons: Hamiltonian evolution and the first law'',
Phys. Rev. D \v{62}, (2000) 104025.

\bibitem{ADM}
R. Arnowitt, S. Deser, C.W. Misner,
``The Dynamics of General Relativity'',
in {\it Gravitation: An Introduction to Current Research}
ed. L. Witten (New York: Wiley) (1962).

\bibitem{noethercharge1}
V. Iyer and R.M. Wald,
``Some Properties of Noether Charge and a Proposal for Dynamical
Black Hole Entropy'',
Phys. Rev. D \v{50}, (1994) 846--864. 

\bibitem{noethercharge2}
R.M. Wald and A. Zoupas,
``A General Definition of ``Conserved Quantities'' in General Relativity
and Other Theories of Gravity'',
Phys. Rev. D \v{61}, (2000) 084027. 

\bibitem{noethercharge3}
V. Iyer and R.M. Wald,
``A comparison of Noether charge and Euclidean methods for Computing
the Entropy of Stationary Black Holes'',
Phys. Rev. D \v{52}, (1995) 4430--4439. 

\bibitem{weylembed}
M. Spivak, 
{\it A comprehensive Introduction to Differential Geometry},
(Publish or Perish) (1999). 

\bibitem{gaussbonnet}
S. Kobayashi and K. Nomizu, 
{\it Foundations of Differential Geometry}, 
(Wiley: New York) (1969). 

\bibitem{unpublished}
S. Anco and K. Wong, 
``Quasilocal energy-momentum from the gravitational action'',
Technical Report (University of British Columbia) (1993);
S. Anco, 
``Einstein's equation with spatial boundary conditions'',
Talk given at University of Oregon, Relativity Seminar (1995)
(http://oregonstate.edu/$\sim$drayt/GRseminar/GRsem95.html)

\bibitem{Witten}
C. Crnkovi\'c and E. Witten,
in {\it Three Hundred Years of Gravitation},
ed. S.W. Hawking and W. Israel
(Cambridge University Press: Cambridge) (1987) p.676--684. 

\bibitem{WaldLee}
J. Lee and R.M. Wald,
``Local symmetries and constraints'',
J. Math. Phys. \v{31}, (1990) 725--743.

\bibitem{Anderson}
I.M. Anderson, 
``Introduction to variational bicomplex'',
in Mathematical Aspects of Classical Field Theory,
Contemp. Math. \v{132} (1992), 51--73.

\bibitem{Wald}
R.M. Wald, 
``On identically closed forms locally constructed from a field'',
J. Math. Phys. \v{31} (1990), 2378--2384. 

\bibitem{Nester2}  
J.M. Nester, ``A Covariant Hamiltonian for Gravity Theories'',
Mod. Phys. Lett. A \v{6}, (1991) 2655--2661;
J.M. Nester, ``Some Progress in Canonical Gravity'', in Directions in
General Relativity,
ed. B.L. Hu, M.P. Ryan and C.V. Vishveshwara 
(Cambridge University Press: Cambridge) Vol I (1993) 245--260.

\bibitem{ONeillII}
B. O'Neill,
{\it Elementary Differential Geometry},
(Academic Press: New York) (1966).

\bibitem{lightconeex}
N. \'O Murchadha, L.B. Szabodos, K.P. Tod,
``A comment on Liu and Yau's positive quasilocal mass'',
gr-qc/0311006. 

\bibitem{Brinkmann}
H.W. Brinkmann,
``On Riemann spaces conformal to Euclidean space'',
Proc. Nat. Acad. Sci. \v{9}, (1923) 1--3.

\bibitem{spinoranalysis}
In preparation, 
S.C. Anco (2006). 

\bibitem{Hayward}
S.A. Hayward,
``General laws of black-hole dynamics",
Phys. Rev. D. \v{49} (1994), 6467--6474.

\bibitem{criteria}
D. Christodoulou and S.-T. Yau,
``Some remarks on the quasilocal mass''
in {\it Mathematics and general relativity}, 
Contemp. Math. (AMS) \v{71} , (1988) 9--14. 

\bibitem{Wald-book2}
R.M. Wald,
{\it Quantum Field Theory in Curved Spacetime and Black Hole Thermodynamics}
(University of Chicago Press: Chicago) (1994).

\bibitem{Huisken}
G. Huisken, 
``Contracting convex hypersurfaces in Riemannian manifolds
by their mean curvature'',
J. Diff. Geom. \v{31}, (1990) 285--299. 

\bibitem{mcurvflow}
X.-P. Zhu,
{\it Lectures on Mean Curvature Flows},
(American Mathematical Society) (2002).

\bibitem{ChoquetBruhat}
Y. Choquet-Bruhat,
``Positive-energy theorems'',
in {\it Relativity, groups, topology II},
ed. B.S. DeWitt and R. Stora
(North Holland: Amsterdam) (1984).



\end{thebibliography}
\end{document}